\documentclass[iop]{emulateapj}
\usepackage{subfigure}
\usepackage{graphicx}
\usepackage{ifthen}
\usepackage{tabularx}

\newcommand{\ppcf}{Plasma Phys. Control. Fusion}
\newcommand{\adndt}{Atom. Data and Nuc. Data Tab.}

\def\rme{{\rm e}}
\def\rms{{\rm s}}
\def\rmp{{\rm p}}
\def\rmd{{\rm d}}
\def\rmf{{\rm f}}
\def\rmg{{\rm g}}

\shorttitle{SASAL for X-ray and EUV spectroscopy ... }
\shortauthors{Liang et al.}

\begin{document}

\title{X-ray and EUV spectroscopy of various astrophysical and laboratory plasmas \newline
      {\small --- Collisional, photoionization and charge-exchange plasmas}}

\author{G.Y. Liang\altaffilmark{1}, F. Li\altaffilmark{1}, F.L. Wang\altaffilmark{1},Y. Wu\altaffilmark{2}, J.Y. Zhong\altaffilmark{1}, G. Zhao\altaffilmark{1}}

\altaffiltext{1}{Key Laboratory of Optical Astronomy, National
Astronomical Observatories, CAS, Beijing, China}
\email{gyliang@bao.ac.cn}
\altaffiltext{2}{Beijing Institute of
applied Physics and Computational Mathematics, Haidian, Beijing,
China}

\begin{abstract}
Several laboratory facilities were used to benchmark theoretical
spectral models those extensively used by astronomical
communities. However there are still many differences between
astrophysical environments and laboratory miniatures that can be
archived. Here we setup a spectral analysis system for
astrophysical and laboratory ({\sc sasal}) plasmas to make a
bridge between them, and investigate the effects from non-thermal
electrons, contribution from metastable level-population on level
populations and charge stage distribution for coronal-like,
photoionized, and geocoronal plasmas. Test applications to
laboratory measurement (i.e. EBIT plasma) and astrophysical
observation (i.e. Comet, Cygnus X-3) are presented. Time evolution
of charge stage and level population are also explored for
collisional and photoionized plasmas.
\end{abstract}



\section{Introduction}
Since the launch of new-generation X-ray and EUV missions (e.g.
Chandra, XMM-Newton, as well as Hinode and SDO for solar physics),
a large amount of high quality spectra with high-resolution and
imaging have post deep insights for our understanding to universe
objects, including their emission measure, physical environment,
space structure or morphology, heating mechanism and so
on~\citep{GN09}. Moreover various spectral models were constructed
for the understanding of the observational data, such as
Chianti~\citep{LZY12}, MEKAL~\citep{MKL95}, AtomDB~\citep{SBL01},
Cloudy~\citep{FKV98}, Xstar~\citep{KB01} and so on. However, they
are strongly depend on the data accuracy of underlying atomic
processes. So a different branch
--- Laboratory Astrophysics (LA), of astrophysical research appears to
benchmark these theoretical models of celestial
emissions~\citep{FHv04,FFP01} or simulate the astrophysical
phenomenon in morphology directly based upon some scaling method,
such as jets, shocks and magnetic reconnection~\citep{ZLW10} etc,
see the review by \cite{RDR06} for details.

Electron beam ion trap (EBIT) was usually used to benchmark
various spectral models for electron-collision plasmas or help
line identification for coronal-like plasmas due to its
characteristics of dissociation of various atomic processes and of
consistent plasma condition (via the electron density ranging
10$^9$---10$^{13}$~cm$^{-3}$) to astrophysical
cases~\citep{LBC09}. Yet, EBIT runs generally at monoenergetic
electron beams, which differs from astrophysical cases with
thermal electrons. Furthermore polarization effects from
unidirectional beams play an important role on line intensity of
some features. The Livermore EBIT group used this platform further
to simulate charge-exchange produced X-ray emissions in
comets~\citep{BBB03}. However the kinetic energy (tens of km/s) of
trapped ions in EBIT is significantly lower than highly charged
ions in solar wind with velocities of 300--800~km~s$^{-1}$.
Recently, other laboratory platforms were used to modelling the
condition of black-hole objects, such as $Z$-pinch and intense
laser~\citep{FTY09}. However, there are still some gaps between
the laboratory miniatures and astrophysical cases, such as
temperature, density, gradient of temperature/density, equilibrium
status and so on. A self-consistent and complete model is
necessary to make a bridge for the laboratory and astrophysical
plasmas.

In this work, we present a description of an analysis package --
Spectral Analysis System for Astrophysical and Laboratory plasmas
({\sc sasal}) -- for the spectroscopic measurements in laboratory
and their application to astrophysical observations. The theory
and atomic data those incorporated into this model are outlined in
Sect.~\ref{sect_theory}. Sect.~\ref{sect_app} illustrates its
applications for the spectroscopy of plasmas dominated by electron
collision with thermal and mono-energetic (e.g. EBIT case)
electrons and by photoionization. Effects from non-equilibrium and
metastable population on level population and/or charge stage
distribution are examined for electron-collision and photoionized
plasmas. Moreover, the application of charge-exchange X-ray
spectroscopy in comets is presented in this section. The last
section gives a summary and conclusion.

\section{Theory and atomic data } \label{sect_theory}
Optically thin assumption is adopted in this model. The atomic
model used to describe line emission from a particular ionic
species ($X^{q+}$) includes the upper level population due to
electron/proton/photon impact (de-)excitation, electron/photon
ionization from lower neighbor ion ($X^{(q-1)+}$),
dielectronic/radiative recombination (DR/RR) from higher neighbor
ion ($X^{(q+1)+}$), charge transfer in collisions with neutral
atom and molecular, and subsequence radiative decays, either
directly to the ground and lower excited states or via cascades,
as following formula,
\begin{eqnarray}
\frac{d}{dt}n_i^{q+} & = & n_{\rm e}\sum_{j\ne
i}n_j^{q+}(Q_{ji}(T_{\rme}) + P_{ji}(T_{\rme}) ) +
\sum_{j>i}n_j^{q+}A_{ji} \\
 & - & n_i^{q+}\left[n_{\rm e}\sum_{j\ne i}\left(Q_{ij}(T_{\rme})+ P_{ij}(T_{\rme})\right) +\sum_{i<j}A_{ij} \right]
  \\
 & + & n_{\rm e}\left[\sum_{i'=0}^{m^{q-1}}n_{i'}^{(q-1)+}S_{i'i}(T_{\rme}) +
 \sum_{j'=0}^{m^{q+1}}n_{j'}^{(q+1)+}\alpha_{j'i}(T_{\rme})\right]
  \\
 & - & n_{\rm e}\left[\sum_{i'=0}^{m^q}n_i^{q+}S_{ii'}(T_{\rme}) +
 \sum_{j'=0}^{m^q}n_i^{q+}\alpha_{ij'}(T_{\rme})\right] \\
& + & \gamma\left[\sum_{i'}n_{i'}^{(q-1)+}\theta_{i'i}(E)
 - \sum_{j'}n_i^{q+}\theta_{ij'}(E)\right] \\
& + & n_{\rm mol}\left[\sum_{i'}n_{i'}^{(q+1)+}C_{i'i}(T_{\rme}) -
\sum_{j'}n_i^{q+}C_{ij'} \right]
\end{eqnarray}
where $n_i^{q+}$ is the number density of $q+$ charged ions at
$i^{\rm th}$ level state, while $n_e$, $\gamma$ and $n_{\rm mol}$
correspond to the number density of electrons, photons and neutral
atoms/molecoles, respectively. $Q_{ij}$ refers to the
electron/proton impact (de-)excitation rate coefficient, $P_{ij}$
corresponds to the photon (de-)excitation rate coefficient,
$S_{i'i}$ and $\alpha_{j'i}$ correspond to the electron impact
ionization and dielectronic plus radiative recombination rate
coefficients, respectively. $\theta_{ij}$ and $C_{ij}$ are
photoionization and charge-exchange recombination rate
coefficients, respectively. The first, second and third terms in
the right part of the equation correspond to contributions from
electron/proton excitations and subsequent radiative decays. The
terms of Eq.-3 and Eq.-4 denote the contributions from electron
impact ionization and dielectronic plus radiative recombination.
The Eq.-5 and Eq.-6 terms refer to population and depopulation due
to photoionization and charge-exchange from neighbour ions.
Furthermore, transitions among tens or hundreds of levels for each
ion, and ionizations/recombinations to/from several/tens of levels
of the neighbour ions have been assumed by consideration of data
availability and practical applications. Above complex equation
can be simplified to be
\begin{eqnarray} \label{eq_matx}
d\frac{\overrightarrow{N}}{dt} & = & {\bf A}\overrightarrow{N}
\end{eqnarray}
where ${\bf A}$ is a stiff matrix being consist of parameters of
various atomic processes mentioned above.

\subsection{Level energies,radiative decay rates and excitations}
Chianti v7~\citep{LZY12} incorporates a large amount of data for
ions with nuclear number from H to Zn from published atomic data,
which is regarded as the most accurate, widespread and complete
spectral model presently in wavelength range of 1---2000\AA\, by
the solar/stellar community. So Chianti (v7) database is the
baseline data for the present model. For He-like~\citep{WBB01},
Li-like~\citep{LB11}, B-like~\citep{LBZ12}, F-like~\citep{WWB07},
Ne-like~\citep{LB10} and Na-like~\citep{LWB09a,LWB09b}
iso-electronic sequences from single ionized through up to krypton
ions, as well as some astrophysical interested ions
(Si~IX---Si~XIII~\citep{LBZ11,LLB13} and Fe~XIV~\citep{LBC10}),
accurate calculations have been done within intermediate
close-coupling framework transformation (ICFT) or {\sc darc}
$R$-matrix methods under the UK Rmax and APAP
network\footnote{http://www.apap-network.org \label{ft_apap}}, as
well as LA project in China. So these data were used to update the
level energies, radiative decay rates and impact excitations data
of charged H---Zn ions available from Chinati v7. Here, those
theoretical wavelengths were adjusted by using available NIST
v3\footnote{http://www.nist.gov/pml/data/asd.cfm \label{ft_nist}}
level energies for some ions, e.g. N$^{5+}$, O$^{5+}$,
Si$^{3+,4+,8+,...,12+}$, Ar$^{7+,8+,13+,...,15+}$, Ca
Fe$^{16+,21+,23+}$, and Kr$^{26+,31+,33+}$. An extension for ions
up to krypton has been done for above mentioned iso-electronic
sequences in this model. Moreover, the original effective
collision strengthes over large temperature range were used in
present {\sc sasal}, not scaled ones as done in Chianti model.
This benefits the {\sc sasal} data update by direct replacement
from data producers. Users do not need to do an onerous data
scaling. Yet the scaling procedure in Chianti extends its
application to more extensive temperature range than the present
{\sc sasal} that covers limited temperatures available in the
original effective collision strengthes provided by the data
producers. The storage of collision strengths in {\sc sasal}
overcomes this limitation, that will be discussed again later.
For He-like ions~\citep{WBB01}, 49 fine-structure (FS) level
energies from 1s$nl$ ($n=1-5, l\in \rms, \rmp, \rmd, \rmf, \rmg$)
configurations, radiative decay rates and impact excitations
amongst these levels were incorporated into the {\sc sasal} model.
For Li-like ions~\citep{LB11}, valence- and core-electron
excitations up to the $1\rms^25l$ and $1\rms2l4l'$ levels (total
204 fine-structure levels) were included. For B-like
ions~\citep{LBZ12}, 204 close-coupling levels of the
$2\rms^x2\rmp^y (x + y = 3)$, $2\rms^2\{3, 4\}l$, $2\rms2\rmp\{3,
4\}l$, and $2\rmp^23l$ configurations were included, and further
radiative decay rates as well as impact excitation amongst them.
For F-like ions~\citep{WWB07}, 195-FS levels from
$2\rms^22\rmp^5$, $2\rms2\rmp^6$, $2\rms^22\rmp^4\{3,4\}l$ and
$2\rms2\rmp^53l$ configurations, radiative decay rates and impact
excitations amongst these levels were included. For Ne-like
ions~\citep{LB10}, we included the data for 209-FS levels
belonging to $[1\rms^2]2\rms^22\rmp^6$,
$2\rms^22\rmp^5\{3,4,5\}l$, $2\rms2\rmp^6\{3,4,5\}l$ ($l\in \rms,
\rmp, \rmd, \rmf$, and g), and $2\rms^22\rmp^5{6,7}l'$ ($l'\in
\rms, \rmp$, and d) configurations. For Na-like
ions~\citep{LWB09a,LWB09b}, the data for 161-FS levels from
$[2\rms^2]2\rmp^6\{3,4,5,6\}l$, $2\rmp^53\rms3l$ ($l\in \rms,
\rmp, \rmd$), $2\rmp^53\rmp^2$ and $2\rmp^53\rmp3\rmd$
configurations, were included. For some astrophysical abundant
ions (Si$^{8+}$---Si$^{12+}$~\citep{LBZ11,LLB13} and
Fe$^{13+}$~\citep{LBC10}), those data presented in literatures
were incorporated into the {\sc sasal} database.

In order to analyze spectra due to excitations of non-thermal
electrons, i.e. mono-energetic electrons in EBIT plasma, we store
the original collision strength (see references above mentioned)
as well, besides of effective collision strength for thermal
electrons. But its large disk occupation makes the {\sc sasal}
model to be not feasibility for distribution because a large
amount of data within $R$-matrix framework have been incorporated.
Then, we set it to be offline data with multiple averaging codes,
e.g. Gaussian averaging for mono-energetic electrons and
Maxwellian averaging for thermal electrons.

\subsection{Dielectronic and radiative recombinations}
State-of-the-art calculations for dielectronic and radiative
recombination data have been performed by Badnell and
coauthors~\citep{BOS03,Bad06} by using {\sc
autostructure}~\citep{Bad86} for K-shell~\citep{Bad06b,BB06},
L-shell~\citep{CPB04,CPW03,AYB04,ZGK04,MB04,ZGK03,ZGF06,ZGK04b},
Na-like~\citep{AYB06}, Mg-like~\citep{AYY07} and
Al-like~\citep{ANG12} iso-electronic sequence ions from H through
Zn~\footnote{http://amdpp.phys.strath.ac.uk/tamoc/DATA/
\label{ft_amdpp}}, which were incorporated into the present model,
including analysis fits for metastable levels and partial
level-resolved recombination rates. For the partial level-resolved
recombination rates, an automatic level-matching procedure was
used to setup these data and match the level index to that of
above mentioned level energies, where configuration, total angular
momentum $J$, and energy ordering are taken to be good quantum
number~\citep{LWB09b}. These level-resolved dielectronic and
radiative recombination rates significantly benefit the estimation
of recombination emission lines in photoionization plasmas, which
are stored as a separate data file for each ion. If the level
numbers (IDs$_{q-1}$) of recombined ion $X^{q-1}$ stated in
Sect.2.1, is less than the level index of recombined ion in the
partial level-resolved data file, those data recombined to levels
above IDs$_{q-1}$ are discarded directly by consideration of lower
population above one hundred or hundreds of levels. Such
simplified treatment might underestimate recombination line
emission for a specified transition due to cascades from higher
levels. Their total rates are compiled separately  by analysis
fits for calculations of charge stage distribution. For other
M-shell ions, the available data from Chianti v7~\citep{LZY12}
were adopted here.

For He-like ions, dielectronic recombination satellite lines were
implemented by using the following formulae~\citep{Gab72,OP04}
\begin{eqnarray}
I_s^{\rm DR} & = & 4\pi^{3/2}a_0^{3/2}X_{\rm
He}n_{\rme}T_{\rme}^{-3/2}{\rm exp}\left(\frac{-E_s}{kT_{\rme}}
\right) g_sA^r B^a,
\end{eqnarray}
where $X_{\rm He}$ is the ionic fraction of He-like ions in
equilibrium and/or non-equilibrium, $B^a$ is total autoionization
branching ratio $\frac{\Sigma A^a}{\Sigma A^a + \Sigma A^r}$,
$A^r$ is the radiative decay rate, $a_0$ is bohr radius,
$T_{\rme}$ is plasma temperature, $E_s$ and $g_s$ are transition
energy and statistical weight of upper level of a given satellite
line, respectively. Those involved $E_s$, $A^r$ and $ A^a$ values
are generated by online FAC calculation~\citep{Gu08} for Li-like
ions via {\it calc\_fac.pro} program with atomic model of [Li:]
$1\rms^2[2/3/4]l$, $1s2\rms[2/3/4]l$, $1s2\rmp^2$, $1s2\rmp[3/4]l$
and [He:] $1\rms^2$, $1\rms[2/3/4]l$, $2\rms^2$, $2\rms2\rmp$,
$2\rmp^2$ configurations, where $1\rms^22l$ configurations are
used for optimization. For the dielectronic satellite
spectra, \cite{VS78,VS80} generated a large amount of data of the
wavelengths, radiative transition probabilities, and
autoionization rates for He-like ions with Z=4--34. Because no
electronic data available for their calculations, these data are
not compiled into the {\sc sasal} at present, but it is in plan.

Radiative recombination continuum (RRC) is an important feature in
high-resolution spectroscopy of photoionized winds, i.e. Cygnus
X-3~\citep{PCS00}. So we implemented such emissions into the
present {\sc sasal} model. The RRC emissivity by this process is
given as \citep{TG66},
\begin{eqnarray}
\frac{E}{dt dV d\omega} = \frac{dP}{dVdE_{\gamma}} & = & n_{\rm
e}n^{(q+1)+}E_{\gamma}\sigma^{\rm rec}(E_{\rm e})v_{\rm
e}\frac{f(v_{\rm e})dv_{\rm e}}{dE_{\gamma}},
\end{eqnarray}
where $E_{\gamma}$ is photon energy of recombination radiation,
$\sigma^{\rm rec}$ is recombination cross-section, and $f(v_{\rm
e})$ denotes the distribution of electron velocity in a plasma. By
the Milne relation between photoionization (PI, $\sigma^{\rm PI}$)
and recombination cross-section~\citep{RS77}, as well as
Maxwell-Boltzman distribution $f(v_{\rm e})$ for electron
velocities, the above equation can be written as
\begin{eqnarray}
\frac{dP}{dVdE_{\gamma}} & = & \frac{4\pi}{c^2}(2\pi m_{\rm
e}kT_{\rm e})^{-3/2}n_{\rm e} n^{(q+1)+}E_{\gamma}^3
\frac{g^{q+}}{g^{(q+1)+}}  \nonumber \\  & & {\rm
exp}\left(-\frac{E_{\gamma}-I^{q+}}{kT_{\rm e}}\right)\sigma^{{\rm
PI}, q+}(E_{\gamma}),
\end{eqnarray}
where $g^{q+}$ is statistical weight of $q+$ charged ion. As shown
by above equation, photoionization cross-section is the
fundamental parameter for calculation of RRC emissivity, so we
store high-resolution PI cross-section from data producers (e.g.
Badnell group) for radiation calculations. In accessing the data
of~\cite{VFK96}, we slightly modified `phfit2.f' program to be
accessed by {\sc idl} `spawn' command implicitly to obtain the PI
cross-section at different photon energies $E_{\gamma}$. The
bin-size of photon energies depend on the line width selected for
line emissions before calculation. Here, we adopt the bin-size of
$\Delta E_{\gamma}=\frac{1}{8}\lambda_{\rm fwhm}$. In order to
speed up the calculation for charge stage distribution, especially
for non-equilibrium photoionizing plasmas, we also store the PI
rates with Black-body radiation at 23 temperatures over
10$^{-3}$---1.0~keV. The PI data source will be explained in next
subsection.

\subsection{Photoionization}
For photoionization, the data from the compilation of \cite{VFK96}
were complemented into the present model, who adopted analysis
fits to the nonrelativistic calculations for the ground states of
atoms and ions. Recent partial level-specific photoionization
cross-section$^{\ref{ft_amdpp}}$~\citep{Bad06,WBM09,WGK11,WBG11}
were included systematically by an automatic level matching
procedure in calculations as done for above mentioned
level-resolved recombination rates, to account for contributions
from metastable level population. From H through Zn, the present
model includes the cross sections from multi-configuration
intermediate coupling distorted wave calculations~\citep{Bad06}
for K- and L-shell photoionization as well as Na-like
iso-electronic sequence ions$^{\ref{ft_amdpp}}$. For several
elements, such as Ne, Mg, Si, S, Ar, Ca and Ni, it includes latest
Breit-Pauli $R$-matrix calculations~\citep{WBM09,WGK11,WBG11}. For
a few interested ions (e.g. Carbon ions), $R$-matrix calculation
performed by Nahar and coauthors \citep{NPZ00} were also included
from the website maintained by
them~\footnote{http://www.astronomy.ohio-state.edu/$^{\sim}$nahar/
\label{ft_nahar}}. The cross-section were extended up to infinity
in integration (i.e. Black-body radiation) by a tail from
Krammer's fit of $\sigma_{\rm PI}(E)=\sigma^o_{\rm PI}E_o^3/E^3$.

As mentioned in above subsection for RRC emissivity calculation,
high-resolution PI cross-section from data producers are stored
separately in the present {\sc sasal} model, which can be accessed
in calculations. Although this make the {\sc sasal} model to be
difficult to be distributed to users, either the offline data
access or specified data transfer for these high-resolution PI
cross-section is still possible. The final level-resolved PI
cross-section~\citep{Bad06,WBM09,WGK11,WBG11} including K-shell
vacancy benefits emissivity calculations for K-shell fluorescent
lines as given by~\cite{KPB04},
\begin{eqnarray}
\epsilon & = & E_{\gamma}\int_{E_{\rm
th}}^{\infty}F_{E_{\gamma}}\sigma_{\rm K}(E_{\gamma})\frac{d
E_{\gamma}}{E_{\gamma}}\omega_{\rm K} n N_{X}n_j^{q+},
\end{eqnarray}
where $n$ is the density of radiated plasma, $F_{E_{\gamma}}$ is
local photon flux, $\omega_{\rm K}$ is fluorescence yield, $N_{\rm
X}$ refers to the $X$ elemental abundance and $n_j^{q+}$
corresponds to its level population of photoionizing ion with
charge of $q+$ at $j^{\rm th}$ stage. However, we have not
compiled K-vacancy levels and relevant Auger and fluorescence
yields~\citep{GKK03} for most L-, M- and N-shell ions. So the
present {\sc sasal} model can not be used to analyze fluorescent
lines. The present {\sc sasal} model is also not a self-consistent
solution of level populations and radiative equilibrium for
photoionization plasma, where level populations and ionization
equilibrium are treated separately. In the treatment of the level
population of excited valence levels, only the population by
dielectronic recombination from the ground configuration of
neighbor higher charged ion. Thus it can be used for the analysis
of discrete recombination lines in photoionized plasma. In the
treatment of the ionization equilibrium, total photoinization and
dielectronic plus radiative recombination rates are used. Further
explanation will be given in corresponding subsection of
applications.

\subsection{Collisional ionization}
Total collision ionization cross-section/rate with excitation
autoionization contribution for some ions were given in Chianti
v7~\citep{LZY12}. In order to investigate effects due to
metastable level population, we calculated the level-resolved
ionization cross-section by using FAC~\citep{Gu08} for He-like,
L-shell and Ne-like iso-electronic sequence ions from Li to Zn
elements. The atomic model used for this calculation is given in
Table~\ref{tbl-conf}, where the first configuration group (marked
by underline) is used to obtain the optimal radial potential. In
the compilation and calculation of~\cite{Dere07}, there is no
significant evidence of excitation autoionization (EAI)
contributions for He-like, B-like, C-like, N-like,  O-like and
F-like ions. So we did not include EAI contribution for these
iso-electronic ions. For Li-like ions, the EAI cross sections
include excitations to the $1s2l2l'$, $1s2l3l'$ and $1s2l4l'$
levels.  Here the EAI contributions are considered by using
$\sigma_{\rm EAI}(j)=\sum_k\sigma_{\rm exc}(j-k)B_k^a$ (where
$B_k^a$ is autoionization branching ratio from the $k^{\rm th}$
channel). For Be-like ions, the EAI cross sections include
excitation to the levels of $1s2s^22p$, $1s2s2p^2$, $1s2l^23l'$
and $1s2l^24l'$ configurations. For Ne-like ions, the EAI cross
sections via $2s2p^63l$ and $2s2p^64l$ levels are taken into
account.

\begin{table*}[htb] \hspace{-1.0cm}
\caption[I]{Configurations included in calculations for
level-resolved collisional ionization of
Si$^{4+}$---Si$^{11+}$.}\label{tbl-conf} \centering
\begin{tabular}{lc|c} \hline\hline
Ions & \multicolumn{2}{c}{Configuration~list} \\
     &  q+    &  (q+1)+  \\
\hline
Ne-like & \underline{$2s^22p^6$}, $2s^22p^5[3/4]l$, & $2s^22p^5$, $2s2p^6$, $2s^22p^4[3/4]l$, \\
        & $2s2p^6[3/4]l$               & $2s2p^53l$, $2p^63l$ \\
F-like  & \underline{$2s^x2p^y$} (x+y=7), $2s^22p^4[3/4]l$, & $2s^22p^4$, $2s2p^5$, $2p^6$, \\
        & $2s2p^53l$, $2p^63l$                   & $2s^22p^33l$, $2s2p^43l$, $2p^53l$ \\
O-like  & \underline{$2s^x2p^y$} (x+y=6),           & $2s^22p^3$, $2s2p^4$, $2p^5$,  \\
        & $2s^22p^33l$, $2s2p^43l$, $2p^53l$     & $2s^22p^23l$, $2s2p^33l$, $2p^43l$ \\
N-like  & \underline{$2s^x2p^y$} (x+y=5),          & $2s^22p^2$, $2s2p^3$, $2p^4$ , \\
        & $2s^22p^23l$, $2s2p^33l$, $2p^43l$     & $2s^22p3l$, $2s2p^23l$, $2p^33l$ \\
C-like  & \underline{$2s^x2p^y$} (x+y=4),        & $2s^22p$, $2s2p^2$, $2p^3$,  \\
        & $2s^22p3l$, $2s2p^23l$, $2p^33l$       & $2s^23l$, $2s2p3l$, $2p^23l$ \\
B-like  & \underline{$2s^x2p^y$} (x+y=3), $2s^2[3/4]l$, & $2s^2$, $2s2p$, $2p^2$, \\
        &  $2s2p[3/4]l$, $2p^2[3/4]l$     & $2s[3/4]l$, $2p[3/4]l$, $3s^2$, $3s3p$, $3p^2$ \\
Be-like & \underline{$1s^22s^2$}, $1s^22s2p$, $1s^22p^2$ ,$1s^22s[3/4]l$, & $1s^2[2/3/4]l$, $1s2s^x2p^y$ (x+y=2), \\
        &  $1s^22p[3/4]l$, $1s2s^x2p^y[3/4]l$ (x+y=2)         & $1s2s[3/4]l$, $1s2p[3/4]l$ \\
Li-like & $\underline{1s^2[2}/3/4]l$,$1s2s^x2p^y$ (x+y=2), & $1s^2$, $1s[2/3/4]l$, \\
        & $1s2s[3/4]l$, $1s2p[3/4]l$                     & $2s^x2p^y$ (x+y=2) \\
He-like & \underline{$1s^2$}, $1s[2/3/4]l$, $2s^x2p^y$ (x+y=2) & $1s$, $2l$, $3l$ \\
 \hline
\end{tabular}
\end{table*}

\begin{figure*}[th]
\centering
\includegraphics[angle=0,width=13.0cm]{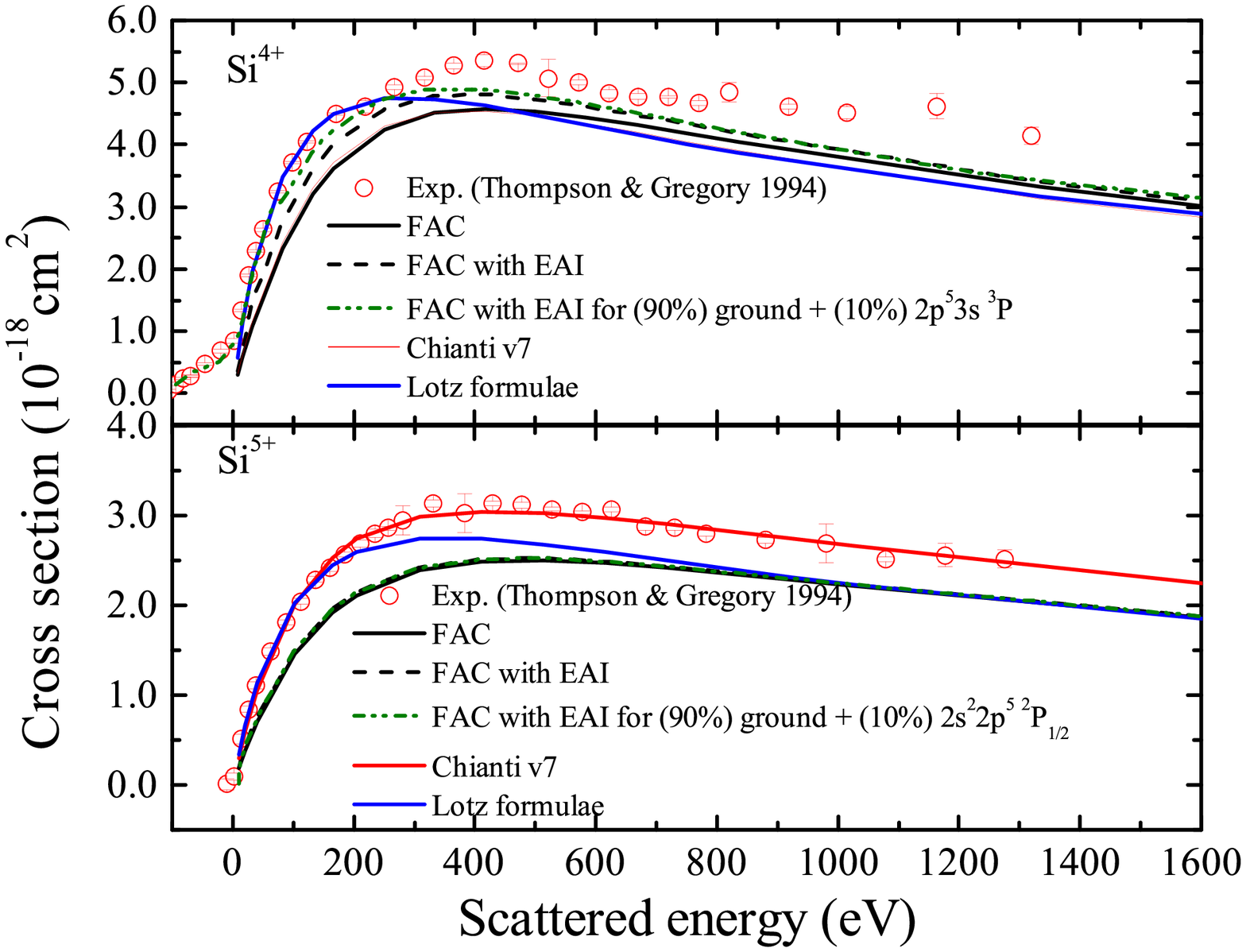}
\includegraphics[angle=0,width=14.0cm]{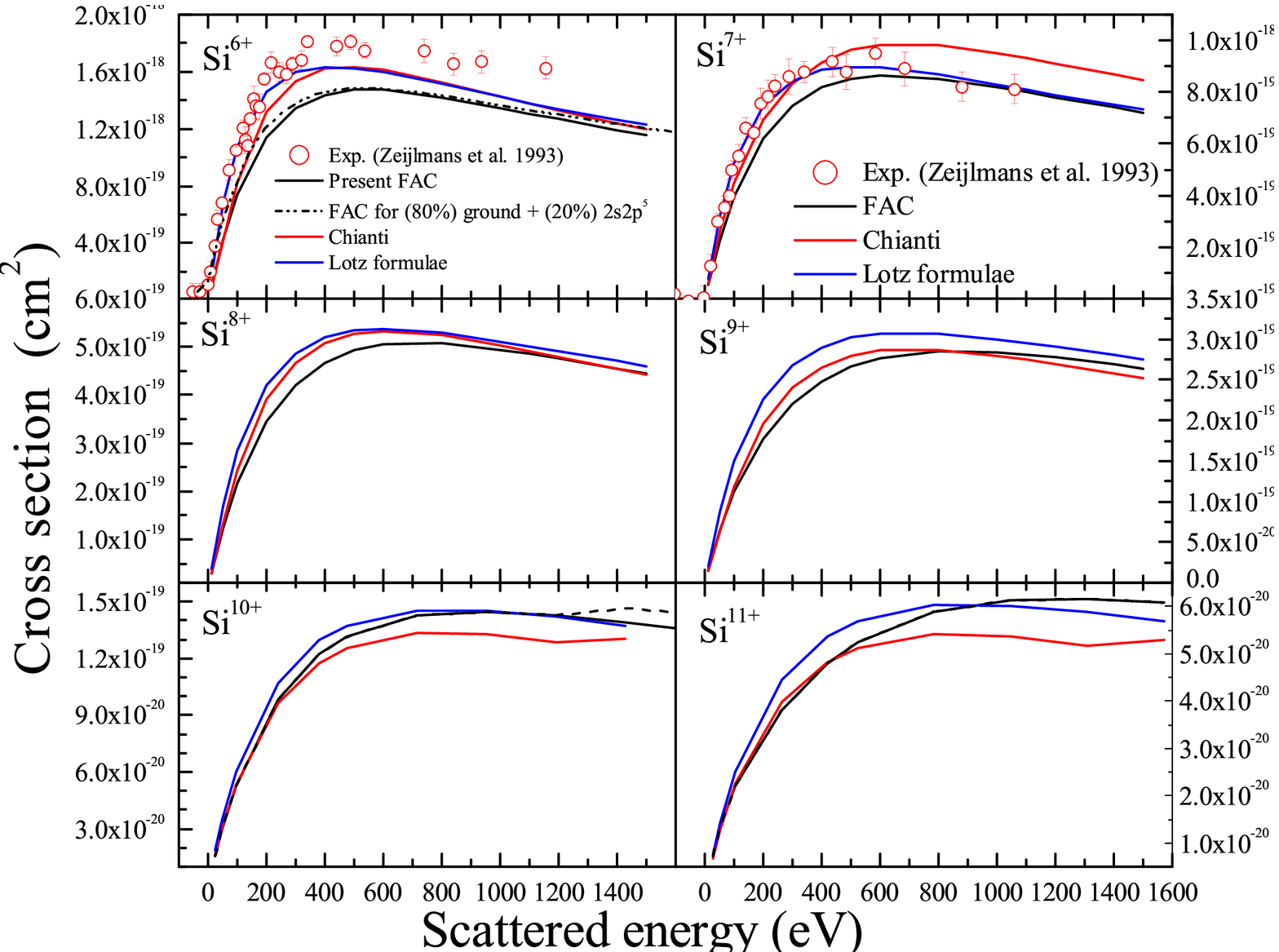}
\caption{\label{fig_ion_cs} Cross-section of electron impact
ionization for Si$^{4+}$
--- Si$^{11+}$ ions at their ground stages from different calculations and database along with available experimental data for Si$^{4+}$--Si$^{7+}$ ions.  [{\em A color version of this figure is
available in online journal}]}
\end{figure*}

Because of recent interest on silicon in laboratory~\citep{FTY09}
and theory~\citep{WSZ11}, as well as its diagnostic application in
astrophysics~\citep{Mil11}, we select highly charged silicon ions
(Si$^{4+}$---Si$^{11+}$) to examine the accuracy of the present
FAC calculation, see Fig.~\ref{fig_ion_cs} for total ionization
cross-section from ground state of each ion. The cross-section
were extended up to infinity by a parameterized formula for
collision strength as given by \cite{ZS90} of $\Omega_{\rm
CI}(E)$=$p_0{\rm ln}u + p_1y^2 + p_2y/u + p_3y/u^2$, where
$u=E/E_{\rm th}$ and $y=1-1/u$. In the following, we will briefly
discuss the results one-by-one.

For Si$^{4+}$ ion, \cite{TG94} concluded the presence of impurity
metastable states to be 5\% by comparing their measured data
at incident electron energies of 50--100~eV with their prediction
from Lotz formula for the direction ionization (DI). However, a
significant discrepancies ($\sim$17\%, see Fig.~2 in \cite{TG94})
between the measured and Lotz cross-section including impurity
metastable contribution were stated, that needs a further
elaborate calculations. The present FAC calculation confirms
that EAI contribution will enhance the ionization cross section
more than $\sim$10--20\% below $I_p$ ($I_p$=167~eV) of scattered
electron energy. When we tentatively assume the impurity
metastable ($2\rmp^53\rms~^3P$) contribution of 10\%, the present
FAC calculation including EAI contribution shows an excellent
agreement with their measurement below the energy of
$\sim2I_p$=334~eV. Above this energy, the difference is also
within 5\% for most reported energies. The difference is less than
1\% between the present DI result and that in Chianti
compilation.

For Si$^{5+}$ ion, fits to measured cross-sections~\citep{TG94}
were used in Chianti v7 compilation. The excitation autoionization
contributions to the cross-section are confirmed again to be
disregarded. Due to the close energy split ($\sim$0.6~eV) for
ground term ${\rm 2s^22p^5~^2P}$, no apparent contribution from
the metastable fine-structure level (${\rm 2s^22p^5~^2P_{1/2}}$)
was observed by \cite{TG94}. By assuming different fractions
of impurity metastable ions, the present FAC calculations
demonstrate that the metastable contribution is negiligible to the
total ionization cross-section, i.e. 10\% impurity fraction
adopted in Fig.~\ref{fig_ion_cs}. 

For Si$^{6+}$, the present FAC calculations are lower than
experimental values by 20\% \citep{ZBG93}, however the Lotz
formula's result and Chianti v7 compilation show better
consistencies with the measurement within 10\%. \cite{ZBG93}
suggested that metastables of $2\rms 2\rmp^5$ configuration may be
responsible for the measured cross section being 20\% larger than
their configuration averaged distorted-wave calculation. But
the present level-resolved FAC calculations demonstrate that the
contribution from $2\rms 2\rmp^5$ levels can be negligible, for
example 20\% metastable fraction used in Fig.~\ref{fig_ion_cs}.

For Si$^{7+}$, the present FAC calculation is lower than Chianti
v7 compilation by 10\%--15\%. Yet it shows a better agreement with
experimental data within 10\% \citep{ZBG93}. For Si$^{8+,9+}$
ions, the present FAC calculations show a good agreement with
Chianti v7 compilation within 10\%.

\begin{figure}[th]
\centering
\includegraphics[angle=0,width=9.0cm]{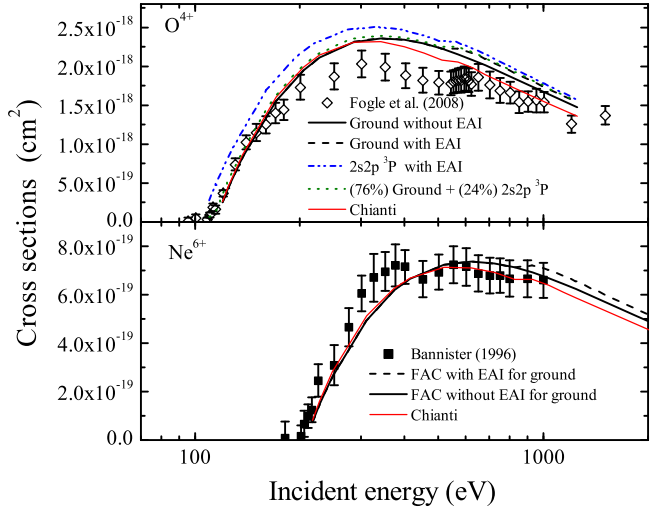}
\caption{\label{fig_ion_cs-o-ne} Comparison of the cross-section
of electron impact ionization between the present FAC calculations
and experimental measurements by \cite{FBB08} and ~\cite{Ban96}
for O$^{4+}$~ and Ne$^{7+}$ ions, respectively. [{\em A color
version of this figure is available in online journal}]}
\end{figure}

For Si$^{10+}$, the present direct ionization cross-section shows
a good agreement with results of \cite{Dere07} within 10\%. When
the EAI contributions due to excitations to $1s2l^3$, $1s2l^23l'$
and $1s2l^24l'$ were included, the resultant total cross-sections
will be enhanced by $\sim$6\% above scattered energy of
$3I_p$ ($I_p=476$~eV). Moreover, the autoionizations via $1s2l^3$
levels are the dominant contribution to this enhancement. We
further check other iso-electronic ions by comparison with
available experimental data, e.g. O$^{4+}$~\citep{FBB08} and
Ne$^{6+}$~\citep{Ban96}~\footnote{http://www-cfadc.phy.ornl.gov/xbeam/cross\_sections.html
\label{ft_cfadc}} as shown in Fig.~\ref{fig_ion_cs-o-ne}. For
O$^{4+}$, the present DI cross-sections are slightly higher than
the experimental data at incident energies of 2$I_p$ --- 8$I_p$
($I_p$=114~eV), but are within 10\%.  At the threshold and high
energy regions, the present results agree with the experimental
measurement within uncertainty. The EAI contribution of
$1\rms$-electron can be noticed above the incident energy of
$\sim$570~eV. By gas attenuation technique, \cite{FBB08} derived a
metastable ($2\rms2\rmp~^3P$) fraction of 0.24$\pm$0.07 in their
experimental ion beam. By assuming the same fraction of metastable
impurity, we also calculate the total ionization cross-section
with the inclusion of EAI. The difference is within 15\% between
the total ionization cross-section and the experimental
measurement~\cite{FBB08}. For Ne$^{6+}$, the EAI contribution is
less than 5\% above $\sim$930~eV. The present DI+EAI calculations
shows a good agreement with the measurement by~\cite{Ban96} within
experimental uncertainty except for a few energies. In the view of
above discussion, the present DI+EAI calculation is reliable for
Si$^{10+}$. 

For Si$^{11+}$, the EAI cross-section is confirmed to be less than
$\sim$1\% of total cross-section. And the present results agree
with Chianti v7 compilation within 10\% below 1 keV. At energies
of 1.0--3.0~keV, the difference between them is about 15\%--20\%.
Above 3.0~keV, the difference becomes smaller again being less
than 15\%. For the simple case of He-like Si$^{12+}$, the present
FAC results agree well with Chianti v7 compilation, and it is not
presented in Fig.~\ref{fig_ion_cs} by consideration of page space.

By above comparison, an uncertainty of 15\% can be accepted
for the present calculations of the collisional ionization. Then
we expect the final accuracy to be within 15\% for the Si
ionization balance.

\subsection{Charge-exchange recombination}
In order to obtain accurate charge exchange cross sections in
ion-atom/molecule (also namely `recipient-donor') collisions, some
sophisticated methods have been developed, including the
molecular-orbital close-coupling method, the atomic-orbital
close-coupling method and time-dependent density theory method.
However, in many cases, the accurate charge transfer cross section
data are very limited due to the difficulties in the sophisticated
treatment of the complex systems, so much simpler multichannel
Landau-Zener (MCLZ) theory offer a flexible choice. Even in
systems for which accurate calculations are possible, application
of the Landau-Zener model can provide useful ``first estimates''
of non-adiabatic transition probabilities. Based on the two-state
Landau-Zener model~\citep{Lan32,Zen32}, the multichannel
Landau-Zener theory with rotational coupling (MCLZRC) have been
developed and was extensively used to estimate the cross-section
of multiply charged ions (`recipient') with hydrogen and helium
(`donor')~\citep{BD80,SO76,JBB83}. In MCLZRC model, electron
transitions happen at the crossing regions $R_n$ of the potential
curve of the collision systems and the transition probability
$p_n$ can be estimated by the Landau-Zener formula, as given by
\begin{eqnarray}
p_n & = & {\rm exp}\left(- \frac{\pi\Delta^2(R)}{2v_R\Delta F(R)}
\right)_{R=R_n},
\end{eqnarray}
where $R_n$ is the curve crossing position, $\Delta R$ is the
energy splitting at the crossing point, $v_{R}=v \sqrt{1-b^2/R^2}$
is the radial velocity wherein $b$ is the impact parameter, and
$\Delta F = (Z-1)/R_n^2$. In the case that the multi-state
coupling dynamics can be reduced to a finite number of two-state
close-coupling problems that are mutually isolated, the
probability $p_n$ of a given exit $n$ will be populated within the
quasi-classical approximation~\citep{SO76}. The original
parameters of $R_n$ and $\Delta R$ can be computed by using
accurate quantum chemical method, for example the multi-reference
singly-doubly excited configuration interaction (MRDCI) method.
However, such quantum chemical methods are too expensive in
computational time. \cite{SO76} have proposed an approximation
estimation method, in which the crossing point can be obtained
based on the ionization energy of the `donor' and excitation
energy of the recipient ion, and the energy splitting $\Delta R$
can be computed using the following analytical formula (within an
accuracy of 17\%)
\begin{eqnarray}
\Delta(R_n)= 18.26\sqrt{Z}{\rm exp}\left(-\frac{1.324
R_n}{\sqrt{Z}}\right).
\end{eqnarray}
Using the parameterized MCLZRC model of \cite{SO76}, the charge
exchange cross section can be computed quickly and we compile this
parameterized MCLZRC code into the present model to estimate
charge exchange (CX) cross-section online for various `recipient'
ions with donors, i.e. hydrogen and helium.

At present version, only the parameterized MCLZRC code was
compiled, and original MCLZRC code is in plan due to complicate
calculation for avoided crossing point and the adiabatic splitting
energy. In this work, we firstly extract level energies of
captured ions from the {\sc sasal} database, then obtain the
averaged energy for each configuration $nl$ or $LS$ term. From
these averaged energies and other necessary parameters (i.e.
captured ion potential, donor potential and polarization, and
exponent of a single orbital wave function), we can derive the
avoided crossing point  $R_n$ in this interaction, and the
adiabatic energy splitting at $R_n$. Furthermore the $nl$-manifold
CX cross-section will be obtained. The resultant $nl$-manifold CX
cross-section is distributed to each level by statistical
weighting. The whole calculation for selected ion can be done
implicitly when this approximation is selected, not compiled data
from published papers or public webisites.

\begin{figure}[th]
\centering
\includegraphics[angle=0,width=9.0cm]{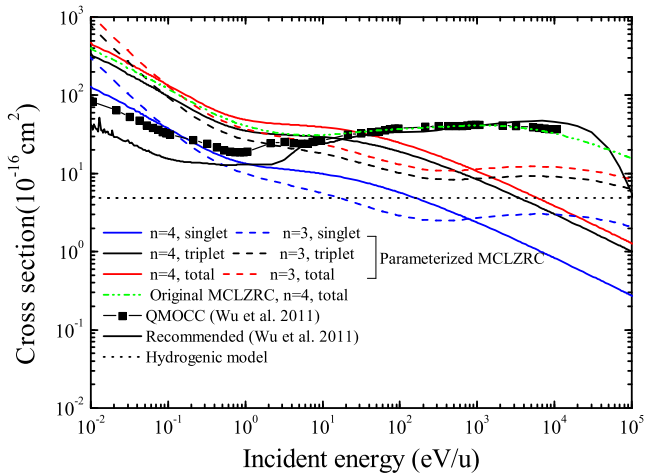}
\caption{\label{fig_cx_cs} Comparison of charge exchange
cross-section of H-like nitrogen (N$^{6+}$) with atomic hydrogen
into the $n$=3 and 4 channels. Filled square and solid curve with
resonances are the recommend data and molecular-orbital
close-coupling calculations from~\cite{WSL11}. Horizontal dotted
line is from the hydrogenic model that adopted by~\cite{WSL97}.
Solid and dotted curves are from the present parameterized MCLZRC
calculation for $n=4$ and 3 channels, respectively. Dashed
double-dot line corresponds to the original MCLZRC calculation for
$n=4$ channel. [{\em A color version of this figure is available
in online journal}]}
\end{figure}

In the following, we present the CX cross-section of N$^{6+}$
colliding with H to check the reliability of this approximation.
Figure~\ref{fig_cx_cs} shows a comparison of the CX cross-section
between the present parameterized multi-channel Landau-Zener
calculations and previous QMOCC calculation~\citep{WSL11} for the
collision of N$^{6+}$ ion with neutral H via $n=3, 4$ channels.
The original MCLZRC calculation ($n=4$) shows an excellent
consistency with recommended data over larger energy region, its
realization of online calculation is in plan. The figure also
demonstrates that the present parameterized MCLZRC calculation is
a acceptable choice for estimation of CX contribution to observed
line emission. For solar wind velocities of $\sim$200--800~km/s
($\sim$200--3500~ev/u), it shows a better agreement between the
parameterized MCLZRC calculation and the recommended data from
\cite{WSL11}. At lower recipient energies $E <1.0$~keV, the donor
electron prefers to transfer to $n=4$ channel of the recipient
ion, but this preference will shift to $n=3$ channel at higher
energies of $E > 1.0$~keV. The widespread of such accuracy of the
parameterized MCLZRC still needs further examination with
elaborated calculations  from QMOCC or CTMC method for other ions.
However such data for astrophysical abundant ions are very scarce.
Anyway, this comparison posts insights for the accuracy of the
parameterized MCLZRC calculation. Moreover energy dependent CX
cross-section can be estimated, better than hydrogenic model, from
which the solar wind dynamics can be estimated.

In collision with other donors, such as hydrogen gas (H$_2$),
water (H$_2$O), carbon monoxide (CO), carbon dioxide (CO$_2$), and
methane (CH$_4$), the polarizability (4.5) and exponent (-1.0) of
single orbital wave function for H was used for the estimation of
CX cross-section.

Furthermore, we complement the hydrogenic model that adopted by
\cite{WSL97} into the present spectroscopic model to estimate the
CX cross-section, that is,
\begin{eqnarray}
\sigma & = & 8.8\times10^{-17}\left(\frac{q-1}{q^2/2n^2 - |I_{\rm
p}|}\right)^2~,
\end{eqnarray}
where $q$ is the charge of recipient ion, $n$ the principal
quantum number of captured ion with peak distribution at $n = q
\sqrt{\frac{1}{2|I_{\rm p}|}}\left(1+\frac{q-1}{\sqrt{2q}} \right
)^{-1/2}$, and $|I_{\rm p}|$ is the ionization potential of donor
in atomic units (i.e., 1~a.u. = 27.2~eV). The $n$-manifold cross
section is distribute into $l$-subshell according to the
distribution function of $W(l,n)=
\frac{(2l+1)[(n-1)\!]^2}{(n+l)\!(n-l-1)\!}$ for low $n\leq8$ and
of $W(l,n)= (\frac{2l+1}{Z}){\rm exp}(-\frac{l(l+1)}{Z})$ for high
$n>8$ values. Finally, the level-resolved CX cross-section are
obtained by the relative statistical weight of each level. The
total $n$=4 CX cross-section 4.89$\times10^{-16}$~cm$^2$ is in
agreement with the recommend data by~\cite{WSL11}, see horizontal
line in Fig.~\ref{fig_cx_cs}.

Additionally, we compiled the CX cross-section from some published
literatures. For example, the electron-capture in collision
between bared oxygen (O$^{8+}$) and hydrogen [H(1s)] performed by
\cite{SGB83}, is compiled into the present model by weighting
$l$-sublevel cross-section with relative statistical weight of
each level, in which the $l$-sublevel cross-sections were derived
from $n$-manifold cross sections $Q_n$ and relative $l$-sublevel
probabilities $P_l$ given in that work. The classical trajectory
Monte Carlo (CTMC) calculation performed by \cite{OOB07,OO08} for
single electron capture of bared and hydrogenic recipients
(carbon, oxygen and neon) with water (H$_2$O) are also compiled
into the present model with same statistic weighting for the
hydrogenic recipients as done by them. The cross-section of
electron transfer in collisions of bared and hydrogenic carbon
with hydrogen gas are from the OPEN-ADAS
database.\footnote{www.open-adas.ac.uk} For the collision between
hydrogen-like nitrogen (N$^{6+}$) and atomic hydrogen,
\cite{WSL11} adopted quantum-mechanical molecular-orbital
close-coupling (QMOCC) method, from which recommended data for LS
term-resolved over low and high energy regions are compiled into
the present model.

In summary, both the parameterized MCLZRC and the hydrogenic
methods will are complemented into the present spectroscopic model
to perform online calculation to obtain the charge-exchange
cross-section.

\section{Applications and discussions} \label{sect_app}
We use the present {\sc sasal} model to analysize the X-ray and/or
extreme-ultraviolet spectroscopy in coronae-like, photoionzed and
geocoronal astrophysical and/or laboratory plasmas.

\subsection{Electron-collision dominant plasma}
For the spectroscopy of thermal equilibrium, the present model is
basically same with that of Chianti v7~\citep{LZY12} for those
ions without data update mentioned in Sect.~2.1 because that
database is the baseline data in the present model. As explained
in Sect.2.1, new atomic structure and excitation data are
incorporated for some ions, which will help new line
identification and will improve spectroscopic diagnostic, see
detail for Fe~XIV spectroscopy in the work of~\cite{LBC10}.
Contributions of recombination and ionizations from/to neighbour
ions can also be explored in the line emissions, in which the same
procedure is adopted as in the discussion of metastable effect in
Sect.3.1.2. Additionally, original collision strengths are stored
as offline data for non-thermal electron distributions due to its
large requirement of disk size, which origins are cited in
Sect.~2.1 for the discussion of excitations. That is, users can
setup any forms of electron energy distribution to avoid the
invalidity problem of detail balance between the excitation rates
and de-excitation rates for non-thermal electrons. In this work,
we will pay attentions on effects from non-thermal electron
(mono-energetic beam, e.g. electron beam ion trap plasma),
metastable population, as well as their time dependence.

\subsubsection{Non-thermal effect}
\begin{figure}[th]
\centering
\includegraphics[angle=0,width=9.0cm]{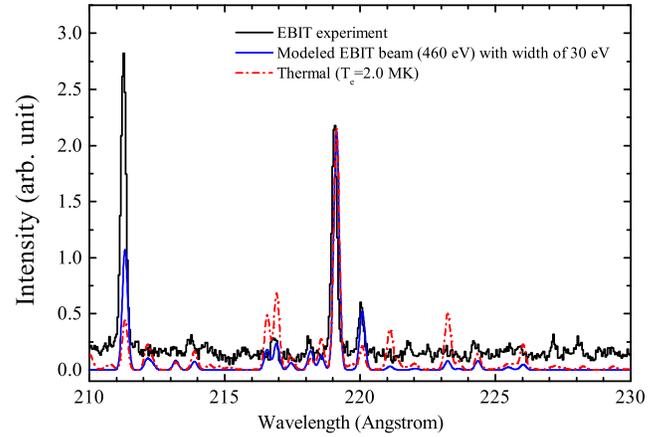}
\caption{\label{fig_mn_th} Theoretical spectra of Fe~XIV at
thermal ($T_{\rm e}$ = 2.0~MK) and modelled mono-energetic
($E_{\rm e}$ = 460~eV) electrons, along with measurement at
Heidelberg EBIT facility~\citep{LBC10}.  [{\em A color version of
this figure is available in online journal}] }
\end{figure}

Electron beam ion trap (EBIT) has been regarded as a better choice
to benchmark various theoretical models for coronal-like plasmas
due to its electron density being consistent with astrophysical
cases~\citep{Bei03}. However, it is usually operated with
monoenergetic electron beams, which will introduce non-negligible
polarization as illustrated by \cite{LBC09}. So a direct
comparison between the EBIT measurement and thermal prediction
maybe have potential problem as done in the work of~\cite{BL12}.
In Fig.~\ref{fig_mn_th}, we demonstrate the theoretical spectra of
Fe~XIV in a thermal plasma with temperature of $T_{\rm e}$=2.0~MK,
and in a modelled mono-energetic beam of 460~eV with beam width of
30~eV at an electron density of 10$^{10}$~cm$^{-3}$, along with
the experimental measurement at Heidelberg EBIT
facility~\citep{LBC10}. For comparison, both the theoretical
spectra are normalized to the experimental one at
$\sim$219.1~\AA\, from the ${\rm 3s^23d~^2D_{5/2}} \to$ ${\rm
3s^23p~^2P_{3/2}}$ transition. For some weak emissions, their line
intensities become lower than those in case of thermal plasma,
while some emissions become stronger. For example, weak emissions
around 221.1~\AA\, and 223.2~\AA\, due to ${\rm 3s3p3d~^2D} \to$
${\rm 3s3p^2~^2P}$ transitions, and lines around 216.6~\AA\, and
216.9~\AA\, due to ${\rm 3s3p3d~^4D} \to$ ${\rm 3s3p^2~^4P}$
transitions, becomes weaker in the modelled mono-energetic beam
than those in thermal plasma. In contrast, ${\rm 3s^23d~^2D_{3/2}}
\to$ ${\rm 3s^23p~^2P_{1/2,3/2}}$ transition lines at 211.3 and
220.1~\AA\, become stronger in mono-energetic case. The large
difference between the measurement and theories is resulted from
low density (10$^{10}$~cm$^{-3}$) adopted here, and its
density-sensitivity as demonstrated by \cite{LBC10}.

\subsubsection{Time evolution of level population and ionic fraction}
During the impulsive phase of solar flare, departures from
ionization equilibrium could occur~\citep{DW13}. During gradual
phase, ionization equilibrium assumption is usually valid due to
higher densities~\citep{BC10}. \cite{Mil11} has pointed out that
correlation between Doppler and nonthermal velocities during
impulsive C-class flare strongly dependent on the ionization
equilibrium assumption. \cite{SH10} clearly shown the minimum and
and maximum timescales to ionization equilibrium for each element
from carbon through up to nickel over temperatures of
10$^4$---10$^9$~K.

\begin{figure}[th]
\centering
\includegraphics[angle=0,width=9.0cm]{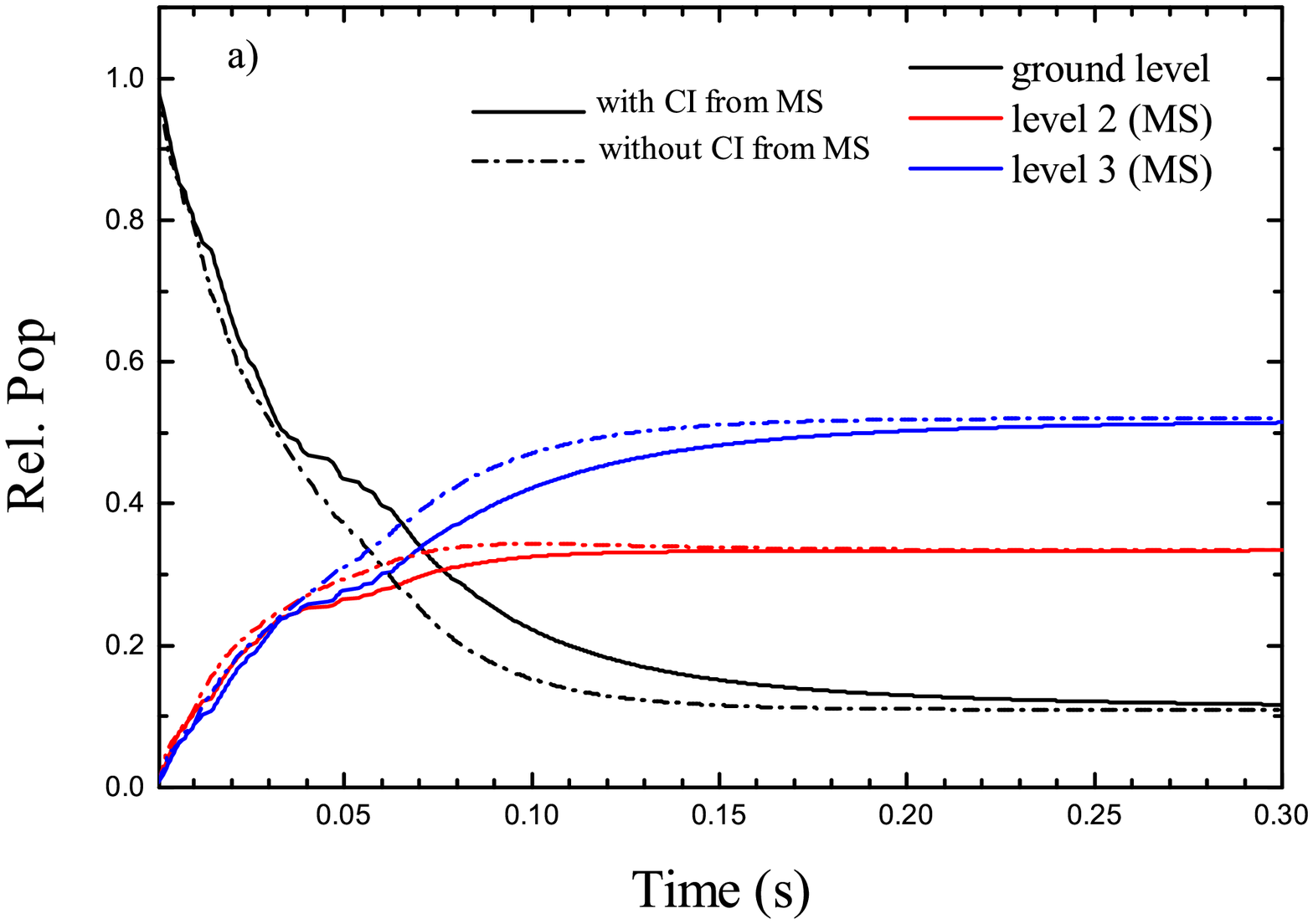}\\
\includegraphics[angle=0,width=9.0cm]{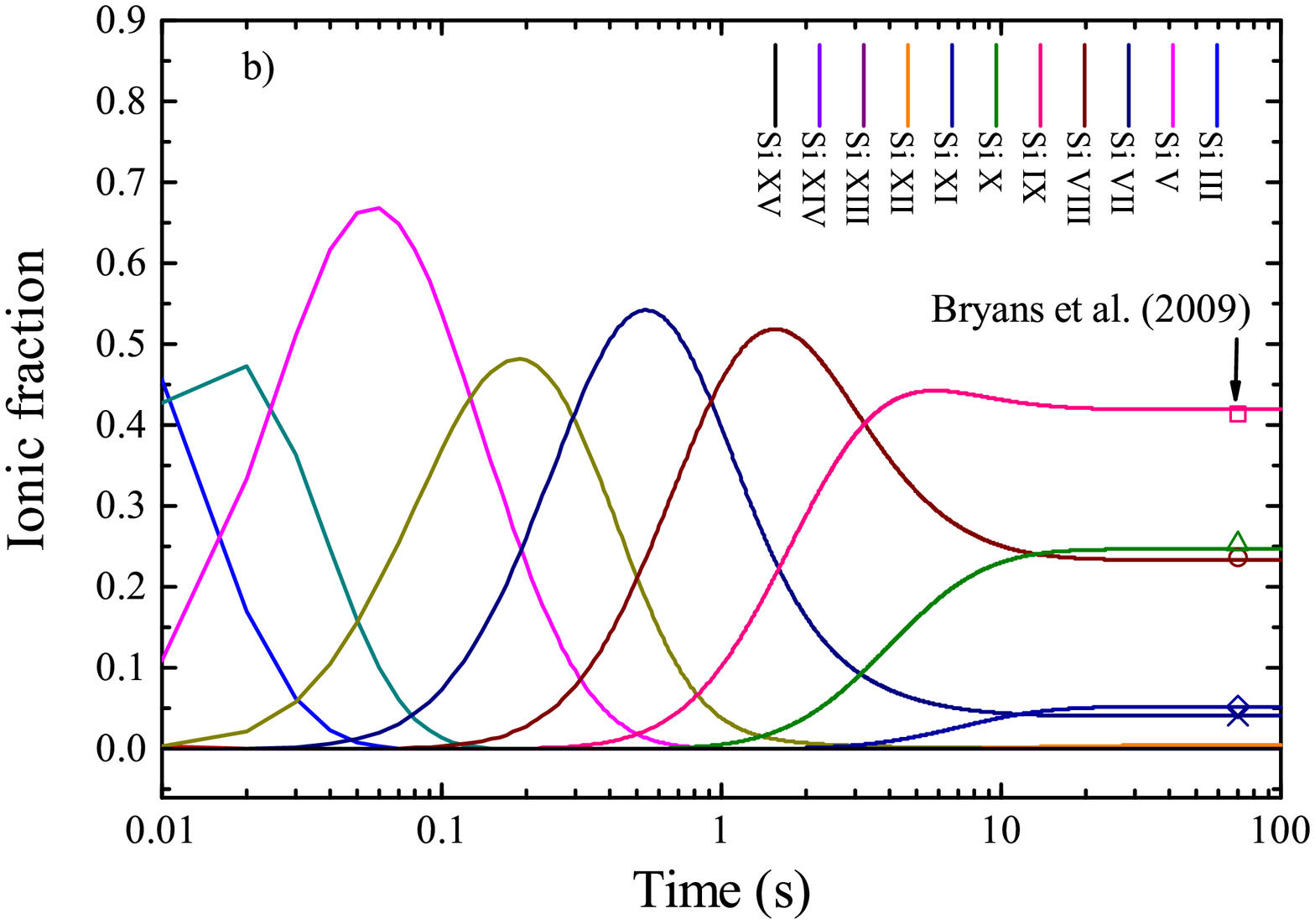}
\caption{\label{fig_lvl_pop} Time evolution of level population
and ionic fraction at an electron density of 10$^{10}$~cm$^{-3}$.
(a) Relative level population of the ground and metastable states
(MS, ${\rm 1s^22s^22p^2~^3P_{0,1,2}}$) of Si~IX with (solid) and
without (dashed-dot) metastable ionization from/to neighbour ions
for a modelled electron beam of 500~eV. (b) The fraction of
various charged silicon as a function of time (s) at a thermal
plasma with temperature of $T_{\rm e}$=1.2~MK. Symbols marked by
an arrow correspond to the values of~\cite{BLS09} at the same
temperature. [{\em A color version of this figure is available in
online journal}]}
\end{figure}

Here by solving time-dependent rate equation, we investigate the
time evolution of level population and ionic fraction.
Fig.~\ref{fig_lvl_pop}-a illustrates the time evolution of level
population for the 3 lowest-lying levels (${\rm
1s^22s^22p^2~^3P_{1,2,3}}$) of Si$^{8+}$ at an electron density of
10$^{10}$~cm$^{-3}$ and a modelled beam energy of 500~eV. Being
less than $n_{\rm e}t = 1.5\times10^9$~cm$^{-3}$s, the ionic level
populations achieve equilibrium, that is within the timescale
($n_{\rm e}t\simeq$1.0$\times10^7$--5.0$\times10^{11}$~cm$^{-3}$s)
of silicon charge stages to achieve equilibrium at $T_{\rm
e}=1.2$~MK that given by \cite{SH10}. Present calculation gives a
time-scale of $n_{\rm e}t\simeq$2.0$\times10^{11}$~cm$^{-3}$s for
silicon at the same thermal temperature with neutral initial
stage, see Fig.~\ref{fig_lvl_pop}-b.  The ionic fraction shows an
excellent agreement with the data of~\cite{BLS09} when the plasma
evolves to equilibrium, see marked symbols in
Fig.~\ref{fig_lvl_pop}-b. The electron density is about
10$^{10}$~cm$^{-3}$, which is a typical value for active regions
before flare event~\citep{BTD00}. So non-equilibrium effect should
be considered when analyzing spectra of solar flares with high
time cadence ($<$10~s, i.e. {\em Solar Dynamics Observatory}).

Effect from metastable populations are also explored on the time
evolution of level population, where two super-levels [L] and [H]
will be constructed to take the contribution of the recombination
and ionization from/to neighbour ions into account, respectively.
Time-dependent ionic fraction is used here. However, an
equilibrium assumption has been adopted for neighbour ions when
obtaining their level population. The product of metastable level
populations and recombinations/ionizations
$\sum_{i'}n_{i'}^{(q-1)+}S_{i'i}$ or
$\sum_{i'}n_{i'}^{(q+1)+}\alpha_{i'i}$ forms the matrix elements
relevant to the two supper-levels [L] and/or [H]. The metastable
populations are found to be slightly delay the time of level
population by $\sim$40--80~ms to achieve equilibrium, see
dashed-dot curves in Fig.~\ref{fig_lvl_pop}-a.

\subsubsection{Effect from metastable level population}
\begin{figure}[th]
\centering
\includegraphics[angle=0,width=9.0cm]{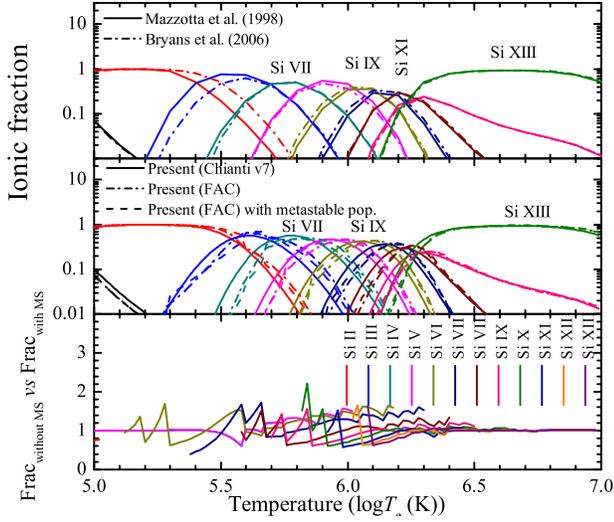}
\caption{\label{fig_ion_bal}Ionization balance for highly charged
silicon ions (III---XIII) between the temperature range log$T_{\rm
e}$~(K)=5--7. {\it Top:} Previous data of \cite{MMC98} and
\cite{BLS09}; {\it Middle:} Present calculations by using the
Chianti v7 data (solid curves), present FAC calculation data for
ionization with (dashed) and without (dashed-dot) metastable
populations at electron density of 10$^{10}$~cm$^{-3}$; {\it
Bottom:} The ratio of the ionic fractions between the calculation
without metastable contributions and that with the metastable
contributions.}
\end{figure}

Present available ionization equilibrium data~\citep{MMC98,BLS09}
that extensively used by astronomical community, are from
calculations at low-density limit, see top panel in
Fig.~\ref{fig_ion_bal}. However, many weak density-sensitive
emission lines are detected, which are usually populated by
excitations from metastable levels. That is metastable population
should play an role on ionic distribution in equilibrium plasma.
Some literatures also attribute observed discrepancies to be
metastable effect. In this work, we investigate this effect on
ionic fraction by using the level resolved ionization and
recombination data mentioned in above section. Here, no line
radiation will be considered, the Eqs. 1--6, will be simplified to
be
\begin{eqnarray} \label{eqn_mslp}
\frac{d}{dt}n_i^{q+} & = &
 n_{\rm e}\left[\sum_{i'=0}^{m^{q-1}}n_{i'}^{(q-1)+}S_{i'i}(T_{\rme}) +
 \sum_{j'=0}^{m^{q+1}}n_{j'}^{(q+1)+}\alpha_{j'i}(T_{\rme})\right]
 \nonumber
  \\
 & - & n_{\rm e}\left[\sum_{i'=0}^{m^q}n_i^{q+}S_{ii'}(T_{\rme}) +
 \sum_{j'=0}^{m^q}n_i^{q+}\alpha_{ij'}(T_{\rme})\right]
\end{eqnarray}

Firstly, we calculate the ionic fraction of silicon in equilibrium
plasma at low-density limit with Chianti v7 database, see solid
curves in bottom panel in Fig.~\ref{fig_ion_bal}, which agrees
well with the data of \cite{BLS09}. That is the present solver for
ionic fraction is correct. By using the present ionization data
from FAC calculation without contributions from metastable
populations (e.g. only the ionization from ground state included),
present ionic fractions (dashed-dot curved in bottom panel of
Fig.~\ref{fig_ion_bal}) slightly shift to higher temperatures.
This small difference can be explained by the small differences in
collisional ionization data as illustrated in
Sect.~\ref{sect_theory}.

Secondly, we calculate the ionic fraction of silicon with
meta-stable contributions at an electron density of
10$^{10}$~cm$^{-3}$, see dashed curves in bottom panel of
Fig.~\ref{fig_ion_bal}. Here, we adopt a different approach from
the Generalized Collisional-Radiative (GCR) method adopted by ADAS
team~\citep{SDO06,LBP09}. There are two separate calculations to
be done in the present calculation. Firstly, we obtain the
meta-stable and/or low-excited level population $n_{i}^{q+,0}$
(hereafter the second upper-script `0' denotes initial ground and
meta-stable populations) of each ion without contribution from
ionization and/or recombination at equilibrium. Then the
ionization $S_{i'i}$ and recombination $\alpha_{j'i}$ rates of
ground and meta-stable levels in Eq.~\ref{eqn_mslp} will be
multiplied by the initial population $n_{i'}^{q+,0}$ or
$n_{j'}^{q+,0}$ for each ion. So we get the ${\bf A}$ matrix in
Eq.-\ref{eq_matx} with matrix elements of
$S_{i'i}n_{i'}^{(q-1)+,0}$ and $\alpha_{j'i}n_{i'}^{(q+1)+,0}$.
The dimension of this matrix dependents on numbers of ground and
meta-stable levels of an iso-nuclear series. For example, there
are $m$ (= \{$m_1, m_2, m_3,\cdots, m_{15} $\} where $m_i$ is the
meta-stable number of $Si_i$ with charge $q+$ of 14-i+1, which is
determined by the maximum level index in ionization and
recombination data files) meta-stable levels for various charged
silicon ions. We will construct a ${\bf A}$ matrix with
(15+$m$)$\times$(15+$m$) dimension. The resultant values
$n_i^{q+}$ in Eq.~\ref{eqn_mslp} are summed to obtain ionic
fraction $\sum_{i=0}^{m_i}n_i^{q+}$ of $q+$ charged ion according
to its meta-stable number $m_i$. Fig.~\ref{fig_ion_bal}
demonstrated that the inclusion of meta-stable populations have an apparent effect on ionic fraction at
$n_{\rme}=10^{10}$~cm$^{-3}$, For some ions, the difference can be
up to a factor of two. Moreover the temperatures of peak abundance
(i.e. Si~VIII) shift to lower temperatures by $\sim$8\% for some
ions.

In comparison to the treatment of GCR method in ADAS package,
the calculations for line emission and ionization distribution are
performed separately in the present {\sc sasal} method. The
first-step procedure in the calculation of ionization distribution
mentioned above, is equivalent to the second terms of the Eq.6--9
in the paper of~\cite{LBP09}, where the effective total
ionizations or recombination rates of ground/metastable levels
include the direct ionizations from ground/metastable levels and
effective excitations to other metastable levels followed by
ionizations to next higher charged ions~\citep{Dic93}. Here, the
effective excitations are realized by solving level-populations
without ionization and recombination contributions from neighbour
ions. This is the main source of the present approximation than
the GCR method. For the line emissions in equilibrium, the present
treatment is similar with that for time-dependent level population
discussed in the above subsection. The contribution from
level-resolved impact ionization and/or recombination of
metastable levels are also included from neighbour two ions. In
principle, the present model partly ignores the cross-coupling
effects in line emissions and ionization distribution. But these
effects is expected to be small for most astrophysical and/or
laboratory plasmas. A qualitative benchmark is still necessary by
a direct comparison with GCR calculation with the same data source
in the near future, that is beyond the scope of the paper.

\subsection{Photoionized plasma}
There are several modelling code for photoionized plasmas, such
as, Cloudy maintained by \cite{FKV98}, Xstar~\citep{KB01}, as well
as PhiCRE constructed by \cite{WSZ11} for laser irradiated
plasmas. For completeness, we also complement the spectroscopic
module for photoionized plasma in the present {\sc sasal} package,
which adopts the data as mentioned in Sect.~\ref{sect_theory}, and
give a brief test below for the present model. At present, only
black-body radiation field is setup and used in this work, while
other radiation fields are routinely in plan and easily setup by
{\sc idl} keywords. As stated in Sect.~\ref{sect_theory},
optically thin assumption was adopted in the present model, then
no radiative transfer is treated in this work. Moreover, an
external constant black-body radiation field is assumed to
investigate the charge state distribution.

\begin{figure}[th]
\centering
\includegraphics[angle=0,width=9.0cm]{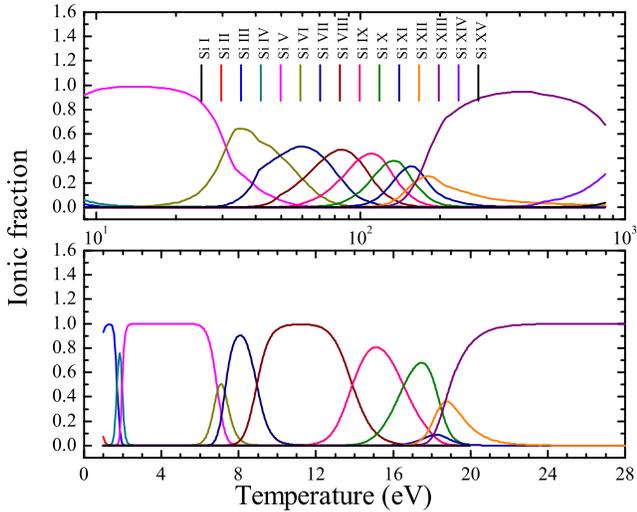}
\caption{\label{fig_ion_bal_phoip} Charge stage distribution of
highly charged silicon ions (I---XV) as a function of thermal
temperature (in eV) and radiation field temperature (in eV) in
collision ({\it top}) and photoionization ({\it bottom}) plasmas,
respectively. [{\em A color version of this figure is available in
online journal}]}
\end{figure}

\begin{figure}[th]
\centering
\includegraphics[angle=0,width=9.0cm]{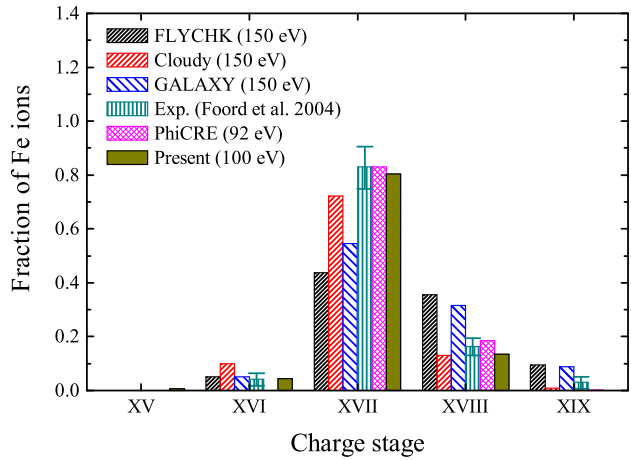}
\caption{\label{fig_ion_frac_pi_fe} Comparisons of the predicted
ionic fraction of iron ions with the laboratory measurement and
previous predictions$^{\ref{ft_sample}}$~\citep{FHv04} from the
codes of {\sc galaxy}, {\sc cloudy}, {\sc flychk} and
PhiCRE~\citep{WSZ11}. [{\em A color version of this figure is
available in online journal}]}
\end{figure}

In calculation of charge stage distribution of photoionization
plasma, only total photoionization and dielectronic/radiative
recombination are included, Eqs.1--6 are simplified to be
\begin{eqnarray}
\frac{d}{dt}n^{q+} & = &
 n_{\rm e}\left[
 n^{(q+1)+}\alpha^{(q+1)+}(T_{\rme})
  - \alpha^{q+}(T_{\rme})\right] \nonumber \\
& + & \gamma\left[n^{(q-1)+}\theta^{(q-1)+}(E)
 - n^{q+}\theta^{q+}(E)\right],
\end{eqnarray}
where $\theta$ and $\alpha$ denote the total photoionization and
recombination rate coefficients, respectively.
Figure~\ref{fig_ion_bal_phoip} (bottom panel) shows the ionic
fraction in photoionized plasma over temperature ($T_{\rm r}$)
0--28~eV of black-body radiation field. For comparison with
collisional plasma, the ionic fraction in collisional equilibrium
is plotted in top panel. For helium-like silicon (Si~XIII), it has
a peak abundance around $T_{\rm e}$=400~eV in collisional plasma,
but it can achieve peak abundance at the radiation field of
$T_{\rm r}$=25~eV. We further use the {\sc sasal} model analyze
the charge state distribution of a photoionized iron plasma
performed by \cite{FHv04} at the Sandia National Laboratory
$Z$-pinch facility. The present result shows a good agreement with
the experimental value at the plasma temperature of 100~eV, as
well as with previous predictions~\footnote{There are small
uncertainty inherent to those data points because they are sampled
from published paper. \label{ft_sample}} from the well-known {\sc
galaxy}, {\sc cloudy} codes at $T_{\rm e}=$150~eV and recent
PhiCRE model at $T_{\rm e}=$92~eV~\citep{WSZ11}, see
Fig.~\ref{fig_ion_frac_pi_fe}. In the present prediction,
experimentally measured electron density of
$2.0\times10^{19}$~cm$^{-3}$ is adopted.

\begin{figure}[th]
\centering
\includegraphics[angle=0,width=9.0cm]{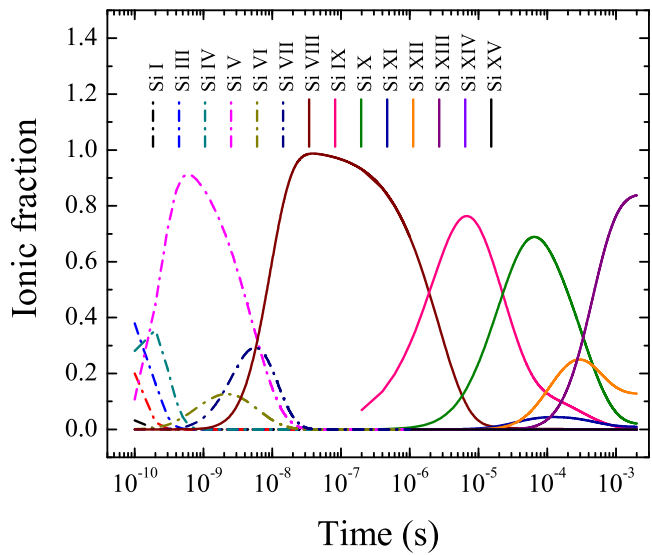}
\caption{\label{fig_ion_bal_phoip_time}  Time evolution of
relative ionic fraction for silicon at an electron density of
10$^{14}$~cm$^{-3}$ and a radiation source of $T_{\rm r}$=30~eV.
[{\em A color version of this figure is available in online
journal}]}
\end{figure}

The time evolution of ionic fraction is also explored for silicon
in photoionized plasma, see Fig.~\ref{fig_ion_bal_phoip_time},
where the electron density is 10$^{14}$cm$^{-3}$ and radiation
temperature is 30~eV. This demonstrates that the maximum
time-scale ($n_{\rm e}t$) is $\sim$2.0$\times10^{11}$~cm$^{-3}$s
to achieve equilibrium at $T_{\rm r}$=30~eV. For high-energy
density plasma ($n_{\rm e}$=10$^{20}$cm$^{-3}$) in laboratory, the
maximum time-scale to achieve equilibrium is
$\sim$6.0$\times10^{11}$~cm$^{-3}$s at $T_{\rm r}$=30~eV. This
time-scale also depends on radiation temperature.

\begin{figure}[th]
\centering
\includegraphics[angle=0,width=9.0cm]{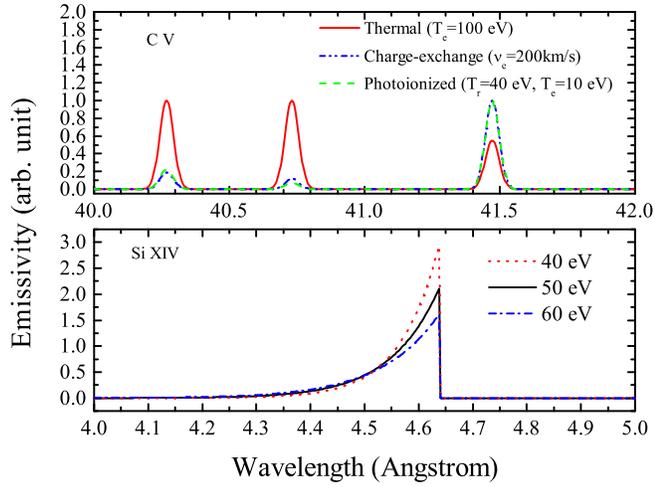}
\caption{\label{fig_c5_line_spec} {\it Top:} Spectroscopy of
He-like carbon ion ($\lambda_{\rm FWHM}=$ 60~m\AA\,) in different
plasmas, such as thermal plasma with electron temperature of
$T_{\rm e}$=100~eV, photoionized plasma with temperature $T_{\rm
r}$=40~eV of Black-body radiation, and  geocoronal plasma due to
the projection of solar wind ion---C$^{5+}$ with velocity of
$v_{\rm e}$=200~km/s into cometary or planetary atmosphere with H
gas. \newline{\it Bottom:} Radiative recombination continuum of
Si~XIV at three different temperatures of 40, 50 and 60~eV.}
\end{figure}

Discrete emissions excited by recombination are important
spectroscopic evidence in a tenuous X-ray photoionized
medium~\citep{PCS00}, presumably the stellar wind from the
Wolf-Rayet companion star~\citep{vCG92}. Cyg X-3 shows a bright,
purely photoionization-driven spectrum and may provide a template
for the study of the spectra of more complex accretion-driven
sources, such as active galactic nuclei. By using the present {\sc
sasal} model, we analyze the discrete recombination emissions of
highly charged carbon by using level-resolved photoionization
cross-section being available from Nahar's
webiste~$^{\ref{ft_nahar}}$. By using Milne relation, we get the
radiative and dielectronic recombination rates.
Figure~\ref{fig_c5_line_spec} shows the spectra of He-like carbon
in photoionized plasma radiated by black-body source with $T_{\rm
r}$=40~eV. For comparison, its spectra in coronal-like plasma with
$T_{\rm e}$=100~eV are also overlap. This clearly demonstrates
that the recombination process favors the population of metastable
level (${\rm 1s2s~^3S_1}$) of the forbidden transition ($f$ line),
that differs completely from the case of collisional plasma, where
resonance transition (${\rm 1s2p~^1P_1} \to$ ${\rm 1s^2~^1S_0}$,
$r$ line) is the strongest one. Such characteristics is usually
used to diagnose the heating mechanism of emitting region in
astrophysical objects. Using the level-resolved dielectronic and
radiative recombination rates compiled in Sect.2, we further
analyze the ratio of $f/(r+i+j)$ ($i,j$ lines are ${\rm
1s2p~^3P_{1,2}} \to$ ${\rm 1s^2~^1S_0}$ transitions) for Si~XIII,
S~XV and Ar~XVII in Chandra observation for Cygnus
X-3~\citep{PCS00}. The resultant ratios are 1.51 (Si~XIII), 1.32
(S~XV) and 0.97 (Ar~XVII) being consistent with Cygnus X-3
observations of 1.3, 1.0, and 0.8, respectively, where we adopt a
temperature of 50~eV estimated by~\cite{PCS00} according to the
shapes of Si~XIV and S~XV RRC features. Estimated temperature
($\sim$60~eV) based on the present RRC emissivity is roughly
consistent with that estimation of ~\cite{PCS00} by an inspection
to Fig.~2 in their work, see bottom panel of
Fig.~\ref{fig_c5_line_spec}.

For completeness, the spectrum due to charge-exchange process is
also overlap in Fig.~\ref{fig_c5_line_spec}, where the projectile
(C~VI) velocity is 200~km/s. Its resultant spectrum is similar
with that in photoionized plasma. The forbidden transition line is
the strongest one. This kind of line formation will be discussed
in below subsection.

\subsection{Geocoronal plasma}
One of the best studied comets is Chandra C/1999 S4 (LINEAR)
observation~\citep{LCD01}, because of its good signal-to-noise
ratio. To discuss our spectral model, we compare our findings with
earlier studies of this comet. Although earlier charge-exchange
(CXE) models~\citep{WSL97,Cra97,HGD01} can explain the observed
x-ray luminosity well, their spectral line shape predictions do
not agree with the observation for the three line ratios of
2.3:4.5:1 at 400, 560, and 670~eV. \cite{BBB03} interpreted the
C/1999 x-ray spectrum by fitting it with their EBIT spectra. That
model takes multiple electron capture into account implicitly, but
the collision energies of 200~eV to 300~eV ($\sim$30~km/s) is far
away from the velocities ($\sim$300--800~km/s) of solar winds.
\cite[see Fig.~3]{BCT07} demonstrated that the hardness ratio has
a strong dependence on the collision velocities below 300~km/s,
implying an overestimation for higher order transition lines than
$n=2 \to$ 1 transition in the work of \cite{BBB03}. However, an
unexpected high C~V fluxes, or low C~VI/C~V ratios were predicted
by \cite{BCT07}. A small contribution from other ions in the
250--300~eV are pointed out by them, that did not included in
their model. \cite{OOB07} included contribution from Mg~IX, Mg~X
and Si~IX ions by using the CTMC intensities of the Balmer
transitions corresponding to a bare projectile with charge equal
to that of the projectile.

\subsubsection{Observation data for LINEAR}
\begin{figure}[th]
\centering
\includegraphics[angle=0,width=9.0cm]{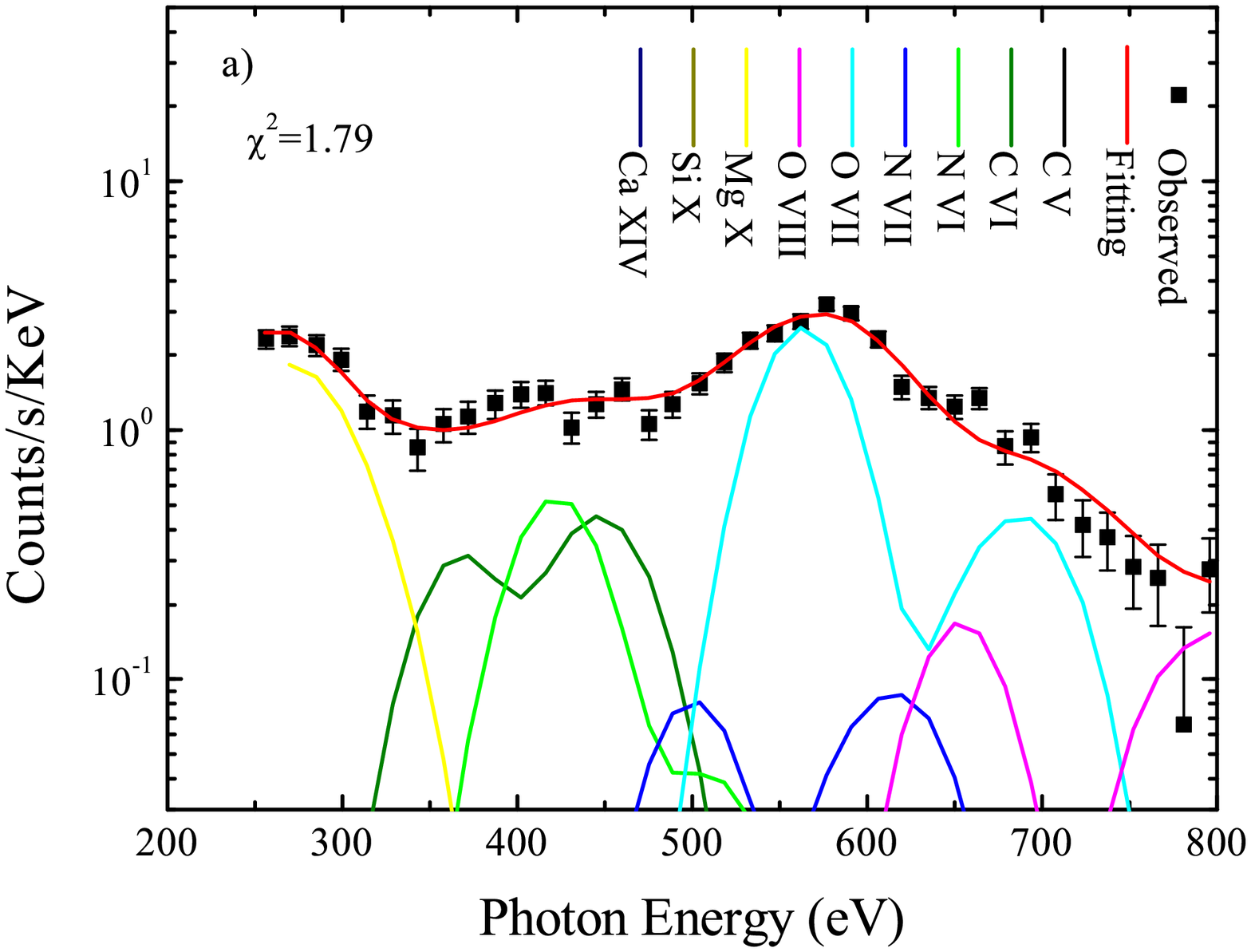}
\includegraphics[angle=0,width=9.0cm]{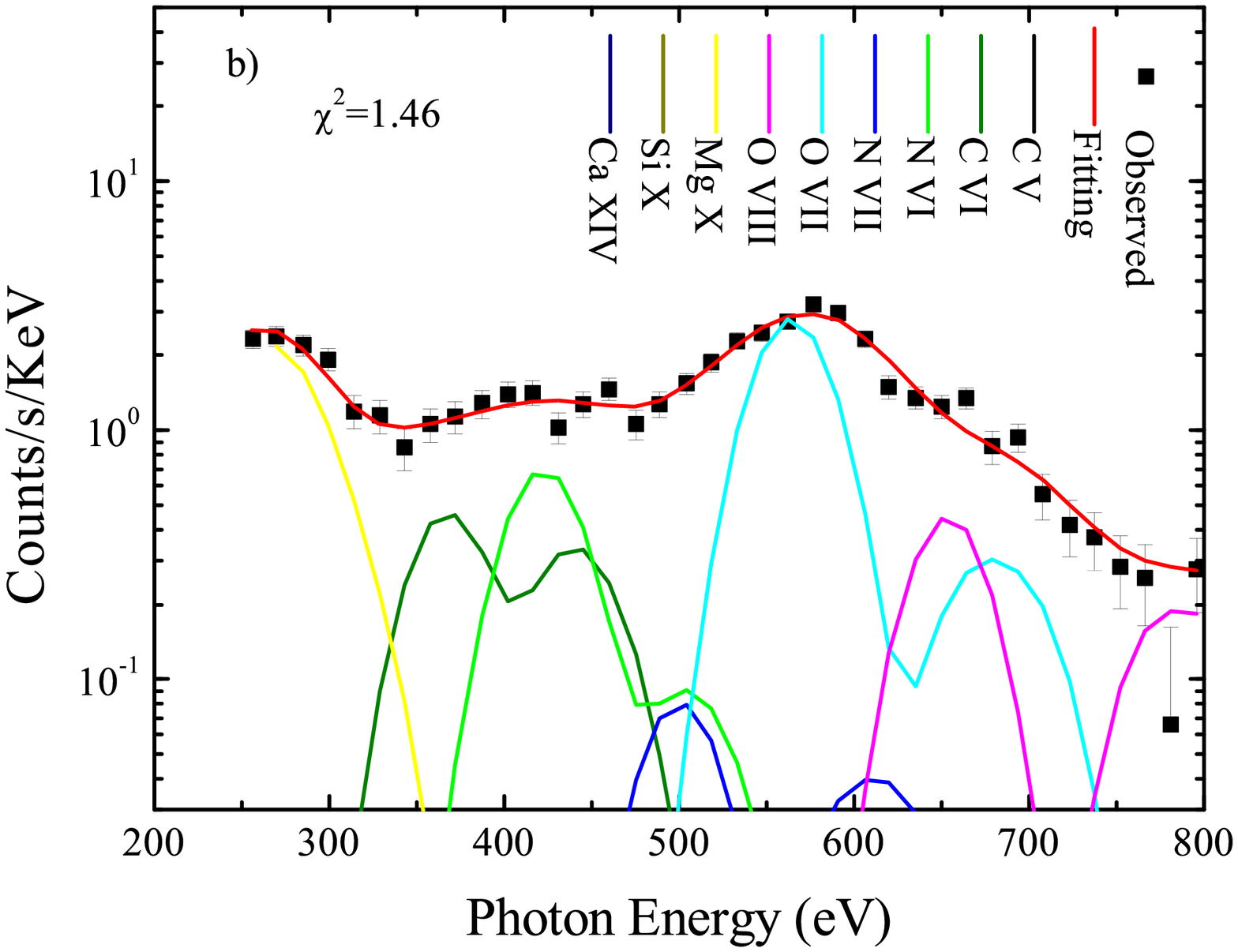}
\caption{\label{fig_linear} Co-added Chandra ACIS observation of
Linear C/1999 S4 before its breakup on 14 July 2000 and its total
fitting performed in {\it Sherpa} with line-width of 50~eV at
different solar wind velocity of 300~km/s (-a) and 600~km/s (-b),
respectively. Specie's contributions are overlapped by multiplying
their CXE model spectra by their fitted fractions. [{\em A color
version of this figure is available in online journal}]}
\end{figure}

The available data set for LINEAR 1999 S4 in {\it Chandra} public
data archive are listed in Table~\ref{table:obscx} before its
breakup. The data reduction is performed by using the Chandra
Interactive Analysis of Observations~(CIAO)
software~(v4.5)\footnote{http://cxc.cfa.harvard.edu/ciao/
\label{ft_ciao}} and by following the science threads for imaging
spectroscopy of solar system objects. In the data reduction, the
source region covers the S3 chip with a circle region with radius
of 4.56$'$, while the background is extracted from S1 chip with a
circle region (radius of 3.78$'$). The data set with Obs$\_$ID of
1748 was not included in the our analysis due to its low
signal-to-noise~(S/N) ratio. Other seven spectra were combined by
using the $combine\_spepctra$ tool to get a high S/N spectra.
Correspondingly, associate auxiliary response files~(ARFs) and
energy dependent sensitivity matrices were obtained by this tool
simultaneously. The resultant observational spectra is presented
by symbols with error bars in Fig.~\ref{fig_linear}. We also
notice the difference of the comet LINEAR observations between the
present extraction and \cite{LCD01}'s data before its breakup as
well as the data adopted by~\cite{OOB07}. This is due to different
source regions adopted in the spectral extraction as illustrated
by~\cite{Tor07}. In the work of~\cite{LCD01}, they adopted EUVE
spatial profiles for the full extent of the comet to correct {\it
Chandra} x-ray flux because the comet overfills the S3 chip and
falls outside its field of view. However, the relative emission
line fluxes have small differences between the present observation
and previous ones, and are within uncertainties. This means the
comparison of relative solar wind abundance with published values
that will be discussed in the following subsection, is still
feasible.

\begin{table}[htb] \hspace{-1.0cm}
\caption[I]{Available observation data sets for LINEAR 1999 S4
before breakup (14 July 2000), with the ACIS-S instrument from
Chandra public data archive.}\label{table:obscx} \centering
\begin{tabular}{lllll} \hline\hline
Obs$\_{\rm ID}$ & Exposure  & Average      & Event & Start  \\
                & Times (ks) & Count Rate & Count &  Time  \\
\hline
584  & 0.95 & 3.00 & 2839 &  04:29:19 \\
1748 & 1.18 & 2.83 & 3331 &  05:06:19 \\
1749 & 1.18 & 2.98 & 3501 &  05:31:29 \\
1750 & 1.19 & 2.85 & 3384 &  05:56:39 \\
1751 & 1.18 & 7.13 & 8383 &  06:21:49 \\
1752 & 1.19 & 6.81 & 8093 &  06:46:59 \\
1753 & 1.18 & 7.46 & 8812 &  07:12:09 \\
1754 & 1.36 & 4.06 & 5516 &  07:37:19 \\
\hline
\end{tabular}
\end{table}

\subsubsection{Fitting in Sherpa with the {\sc sasal} model}
The fitting procedure is done in {\it sherpa} package of CIAO
v4.5.$^{\ref{ft_ciao}}$ A multi-gaussian model are constructed
based on the spectral lines calculated by our {\sc SASAL} package,
as the following formulae,
\begin{eqnarray}
I_{\rm theo}(\gamma) & = & \sum_i A_i\sum_j G_{ij}(\gamma,E_{ij})
\epsilon_{ij},
\end{eqnarray}
where $A_i$ is ionic fraction, $G_{ij}(\gamma, E_{ij})$ is
gaussian profile of a given transition $i\to j$ with the
transition energy of $E_{ij}$ and a given line width, and
$\epsilon_{ij}$ is charge-exchange line emissivity.

In figure~\ref{fig_linear}, we present our CXE fit x-ray spectra
for Linear C/1999 S4 with thousands of lines at velocities of
300~km/s (-a, fast solar wind) and 600~km/s (-b, slow solar wind),
respectively. The collision velocity of 600~km/s adopted here is
consistent with {\sc ace-swepam} and {\sc soho-celias} online data
archive (592~km/s). The line-width is set to be 50~eV, being
consistent with that adopted by~\cite{BCT07} and \cite{LCD01}. It
is narrower than the intrinsic line-width of 110~eV FWHM of the
ACIS-S back-illuminated
CCD~\citep{GBF03}\footnote{http://cxc.harvard.edu/proposer/POG/html/ACIS.html}.
As stated by \cite{LCD01}, it is not significant statistically. In
this model, we include contribution from Mg~X, Si~X and Ca~XIV
charge-exchange emissions. But no Ca$^{14+}$ species can be
estimated in the both solar winds. The inclusion of Mg~X and Si~X
CXE emissions improves the fitting to the spectra between 200~eV
and 300~eV with reduced $\chi^2$=1.46, that region was omitted in
the analysis of \cite{BCT07}. The resultant ionic abundances in
solar winds by this fitting are presented in
Table~\ref{tbl-ionfrac}. For O$^{8+}$, C$^{6+}$ and N$^{6+, 7+}$
ions, the present results are consistent with previous estimations
based upon models. Since there are significant contamination from
C~V and Mg~X emissions below 300~eV, no C$^{5+}$ fraction is
predicted in the fast (600~km/s) solar wind. However, a high
fraction for Mg$^{10+}$ is predicted in the solar wind. Moreover,
the predicted Si$^{10+}$ species (0.10) is consistent to its
fraction (0.12) listed by~\cite{SC00} for the slow wind in their
Table~1, being significantly lower than that listed for the fast
wind by them.  Contributions from unresolved emission lines of
Mg~X and Si~X were not included by \cite{BCT07} and \cite{OOB07}.
And only 10 emissions were included by \cite{Kra06} without
contribution from N$^{6+,7+}$ ions. These can explain the lower
C$^{5+}$ fraction derived according to the present CXE model.
Spectra with much high-resolution, e.g. {\sc astro-h} planned to
be launched in near future, will clarify this problem.

In the fitting with CXE spectra at the collision velocity of
300~km/s, no Si$^{10+}$ and Ca$^{14+}$ ions are derived, and
reduced $\chi^2$=1.79 becomes larger. This indirectly confirms
that the collision energies between recipients and atom/molecular
donors play important roles on spectroscopic analysis for
geocoronal plasmas. Correspondingly, we can estimate the origins
of solar wind ions on solar surface that arrive comets or planet
atmosphere as demonstrated by \cite{BCT07}.

In summary, the complement of multi-channel Landau-Zener method
makes the spectroscopic analysis of charge-exchange plasmas to be
feasible for astronomical community, and confirms it is reliable.
The reliability of hydrogen model is also appropriate, which is
presented by \cite{SFB12}.

\begin{table*}[htb] \hspace{-1.0cm}
\caption[I]{Solar wind abundances relative to O$^{7+}$, obtained
for comet Linear C/1999 S4.}\label{tbl-ionfrac} \centering
\begin{tabular}{lcccccc} \hline\hline
Ion & \multicolumn{2}{c}{This work} & BBB03 & BCT07 &Kra06 & OOB07 \\
    &  600~km/s & 300~km/s &  & & & \\
\hline
O$^{8+}$   & 0.16 & 0.09 & 0.13$\pm$0.03 & 0.32$\pm$0.03  & 0.15$\pm$0.03 & 0.35        \\
C$^{6+}$   & 0.87 & 0.89 & 0.9$\pm$0.3   & 1.4$\pm$0.4    & 0.7$\pm$0.2   & 1.02\\
C$^{5+}$   & 0.00 & 0.02 & 11$\pm$9      & 12$\pm$4.0     & 1.7$\pm$0.7   & 1.05       \\
N$^{7+}$   & 0.07 & 0.06 & 0.06$\pm$0.02 & 0.07$\pm$0.06  &               & 0.03 \\
N$^{6+}$   & 0.59 & 0.14 &  0.5$\pm$0.3   & 0.63$\pm$0.21  &               & 0.29 \\
Mg$^{10+}$ & 4.05 & 3.94 &           &                 &               & 0.97$^e$ \\
Si$^{10+}$ & 0.10 & 0.00 &           &                 &               & 0.80$^e$ \\
 \hline
\end{tabular}
 \flushleft{$^a$~BBB03---\cite{BBB03}; $^b$~BCT07---\cite{BCT07};
 $^c$~Kra06---\cite{Kra06}; $^d$~OOB07---\cite{OOB07}, in which the ion abundance are from ACE-data;
 $^e$~The values of fast solar wind from the work of from~\cite{SC00} that is based on data from
SWICS instrument on {\it Ulysses} solar polar mission. }
\end{table*}

\section{Summary and Conclusions}
In this work, we present a spectroscopic modelling for various
astrophysical and/or laboratory plasmas, including coronal-like,
photoionized, and geocoronal plasmas under optically thin
assumption.

\hspace{0.5cm} (1) For coronal-like plasmas, we construct two
different two modules for thermal and non-thermal (e.g. EBIT)
electrons. Here original collision strengths are compiled into the
database and offline calculation can be done for $R$-matrix
excitation data on local server. This procedure will avoid the
invalidity problem of detailed balance of effective collision
strength between excitations and de-excitations for non-thermal
electrons that extensively adopted by spectroscopic community. So
any form of electron energy distribution can be incorporated into
this module. In modelled mono-energetic case (EBIT), some
emissions are strengthen, while others are weaken than those in
thermal plasma.

\hspace{0.5cm} (2) Effect of metastable population on charge stage
distribution and level population are explored by using
level-resolved ionization and recombination data. It is found to
be small at low-density plasma.

\hspace{0.5cm} (3) Time evolutions of ground and metastable level
population are explored, and found to achieve equilibrium at
$n_{\rm e}t \sim 1.5\times 10^9$~cm$^{-3}$s. This value is
comparable to the timescale ($n_{\rm
e}t\le$2.0$\times10^{11}$~cm$^{-3}$s) of charge stages to achieve
equilibrium. So when plasma departures the equilibrium status, it
means that not only the ionic stages deviate from equilibrium, but
also the level populations possibly departure from equilibrium at
low-density plasma.

\hspace{0.5cm} (4) A module for photoionized plasma with
black-body radiation field is complemented in this model for
charge stage distribution and line emissions based upon fitted
total cross-section, and partial level-resolved  cross-section.
Although the present model is not a self-consistent solution to
most astrophysical photoionized medium, we use it analyze
successfully the discrete recombination emissions and RRC features
in the photoionized wind of Cyg X-3, as well as the charge state
distribution of a laboratory plasma.

\hspace{0.5cm} (5) Charge-exchange spectroscopy of comet--LINEAR
are investigated again here by using the online parameterized
MCLZRC cross-section calculation under the present {\sc sasal}
database. This makes the spectral fitting and analysis to be
feasible for various solar wind ions besides of bared, H-like and
He-like ions. The application to comet--LINEAR reveals that the
present resultant ionic fraction shows a good agreement with
previous ones for O$^{8+}$, C$^{6+}$, N$^{6+}$ and Mg$^{10}$ ions,
wherever a low and high ionic fraction derived for C$^{5+}$ and
N$^{7+}$ than previous ones, respectively. This is due to
inclusion of other emissions and low-resolution of the LINEAR
spectra. High-resolution spectroscopy will clarify this
discrepancy. Moreover, the velocities of solar wind ions have a
significant effect on the determination of ionic fractions.
Additionally, some available $l$- and/or level-distributed
cross-section are compiled into the present {\sc sasal} database.

In conclusion, we setup a self-consistent spectroscopic modelling
package--{\sc sasal} for coronal-like, photoionized and geocoronal
plasmas at equilibrium and non-equilibrium. It is not only a
combination of previous available modelling codes, but also an
extension on metastable effect, time evolution and charge-exchange
dominant plasma etc. Although some assumptions were made, test
applications prove the {\sc sasal} model to be reliable for many
astrophysical and laboratory plasmas. The spectroscopic model for
astrophysical and laboratory cases benefits the community of
laboratory astrophysics due to their inherent differences between
them.

\acknowledgments {We thank the anonymous referee for many
constructive comments on this manuscript, and express our
gratitude to Lijun Gou and Jifeng Liu for their discussions about
Chandra data reduction and fitting. This work was supported by
National Basic Research Program of China (973 Program) under grant
2013CBA01503, and by the One-Hundred-Talents programme of the
Chinese Academy of Sciences (CAS). GYL also acknowledges the
support from the National Natural Science Foundation of China
under grant No. 11273032.}

\clearpage




\end{document}